\documentclass[12pt]{article}
\pdfoutput=1
\usepackage{times}
\usepackage{amsmath}
\usepackage{txfonts} 
\usepackage{geometry}
\usepackage{indentfirst}
\usepackage[utf8]{inputenc}
\usepackage{enumitem,amssymb}
\usepackage{ragged2e}
\newlist{thematic}{itemize}{8}
\setlist[thematic]{label=$\square$}
\usepackage{pifont}
\usepackage{mathrsfs} 
\usepackage{caption}
\usepackage{graphicx, subfig}
\usepackage{float}
\usepackage{booktabs}  
\usepackage{longtable}
\usepackage{threeparttable}

\DeclareUnicodeCharacter{2212}{-}

\begin{document}
\captionsetup[figure]{labelfont={bf},name={Fig.},labelsep=period}
\captionsetup[longtable]{labelfont={bf},name={Table},labelsep=period}
\Large

\begin{center}
\textbf{Measurement of Magnitudes and Effective Temperature with Amateur Telescopes}\\
\normalsize
\hspace*{\fill}\\
Kecheng Qian, Jiaqi Shen\\
\small{United World College Changshu China, No.88 Kun-Cheng-Hu-Xi Road, Changshu, Jiangsu 215500, China}
\newline

\large{\textbf{Abstract}}
\end{center}
\normalsize
In the present study, we developed algorithms that are capable of measuring apparent magnitudes and the effective temperature of stars using raw images shot with amateur telescopes. The regularized Radial Basis Function (RBF) network, one of the machine learning algorithms, was employed to measure the effective temperature, and the simple function fitting method was adopted to measure the apparent magnitude. The achieved results are satisfying. After the white balance and noise cancellation process was simply calibrated, it was demonstrated that the measurements of the effective temperature had mean fraction errors at around 9\%, and the measurements of the magnitudes had absolute error at nearly 0.1.
\newline

$Key\ words:$ apparent magnitude, effective temperature, machine learning,regularized RBF network, amateur telescopes
\newline

\justifying\let\raggedright\justifying
\large
\textbf{1. Introduction}
\newline

\normalsize
By measuring stellar apparent magnitudes and effective temperatures, multiple properties of stars can be more effectively revealed, covering their distance to the earth. In the current digital era, magnitudes are mostly measured using charged couple device (CCD) and professional telescopes in observatories. A commonly employed color index B-V (i.e., the difference between B band magnitude and V band magnitude) to measure the effective temperature of stars is also measured with specialized UBV filters that are mounted on observatory telescopes (Johnson \& Morgan, 1953). 
\newline

Though the measurements of apparent magnitude and color indices have long been mature, most measurements are dependent of either equipment specialized in observatories (e.g., UBV filters) or costly equipment inaccessible to amateur astronomers (e.g., CCDs). The reliance on specialized equipment has compelling reasons, and the reasons are high noise of non-cooled complementary-oxide-semiconductor (CMOS) used in ordinary cameras at high ISO, and their nonlinear response to the intensity of received light (Yadid-Pecht 1999). Thus, under an algorithm used to measure magnitudes and effective temperatures which is of relatively low dependence on data from professional equipment, part of amateur astronomers can facilitate sky surveys. 
\newline

The rest of this paper is organized as follows. In section 2, the amateur telescope and camera, as well as star catalogue applied in this study to acquire labelled data are discussed. In section 3, the procedures of our observation and data reduction are illustrated, which covers white balance calibration, flat field correction, as well as noise cancellation. In section 4, the function fitting employed for magnitudes measurement are analyzed, and its accuracy and compatibility are delved into. In section 5, the ensemble learning algorithm is presented based on the regularized RBF network we adopted to measure the effective temperature. In section 6, the function developed by function fitting is presented, and the accuracy and compatibility of our measurement of the apparent magnitude and the effective temperature are discussed. Lastly, in section 7, all the results of this study are summarized, and the further improvement of the study is suggested  based on the analysis of causes for abnormal errors.
\newline

\large
\textbf{2. Relevant Apparatus and Data Source}
\newline

\textit{2.1. Canon 6Dmk2 and 200mm Cassegrain reflector}
\newline

\normalsize
Canon 6Dmk2 refers to an ordinary digital single lens reflex (DSLR) with a full-frame CMOS. Given that the algorithms applied in this study for measuring the magnitude and the effective temperature is largely dependent of the CMOS adopted, Canon 6Dmk2 was taken as a popular ordinary camera that is used by numerous amateur astronomers. 
\newline

The telescope applied here is a hyperbolic Cassegrain reflector, 200mm in aperture diameter and 1600mm in focal length. For the design of the pure reflector, the telescope exhibits no color aberration theoretically, so it is suitable for this study that highly depends on the color of stars; thus, the critical information about the effective temperature is reflected. As the target stars taken in this study display zenith distance $z<70^\circ$, the effects on the observed magnitude of the stars caused by atmospheric extinction (Karttunen et al. 2017) is defined as$$\delta m=k sec\ z$$Accordingly, the maximum difference in $\delta m$ of two stars with maximum separation in the field of view $dz \approx 1'30''$ at $z<70^\circ$ is assessed as$$d\delta m=\frac{ksinz}{cos^2z}\cdot dz<4\times 10^{-3}k$$As a result, the atmospheric extinction is considered uniform within the small field of view of the telescope. 
\newline

\large{\textit{2.2. Gaia Sky Survey Data}}
\newline

\normalsize
BP magnitude ($m_{BP}$) and effective temperature ($T_{eff}$) from Gaia’s Data Release 2 (Gaia DR2) (Brown et al. 2018) were employed as the expected output. $m_{BP}$ was taken as the expected output of the sum of R,G,B readings in our images since ordinary cameras (e.g., 6Dmk2) has IR infra-red (IR) and ultra-violet (UV) filters; thus, the light could be approximately received within the wave band that is almost identical to that of $m_{BP}$, which is 330-680nm (Brown et al., 2018). Gaia DR2 was selected as it contains data for considerable number of stars between 14$^{th}$ mag. and 16$^{th}$ mag., and the stars we shot for labelled data largely met this range of magnitude. Every star in our images was matched with the star map and then labelled with the corresponding expected output from Gaia DR2. 
\newline

\large
\textbf{3. Observations and Data Reduction}
\newline

\textit{3.1. Observations}
\newline

\normalsize
To shoot the images for labelled data, we went to Dunhuang, Gansu Province, a city near the desert in northwestern China, where the levels of light pollution are relatively low in China. To calibrate the sky-area we were shooting, the stars brighter than 2$^{th}$ mag. were taken as centers of our images. Lastly, stars were used in the two images centering HIP65474 (Spica) and HIP67301 (Alkaid) as our labelled data source. The telescope was pointed to the selected stars and exposed for 30$s$ at 3200 ISO to create images that cover nearly 200 stars brighter than 16$^{th}$ mag. surrounding the central calibration stars. The exposure parameters were adopted to ensure the sufficient number (about 200) of stars in each image as well as the acceptable amount of noise. Given the necessity to take multiple dark fields in astronomical researches, we still decided to leave noise cancellation process to algorithms. As we came to realize that amateur astronomers might not take dark fields when they were observing, the dark field involved into our research might reduce the compatibility of our algorithms. Flat fields were taken because they are unreplaceable by algorithms. 
\newline

\large
\textit{3.2. Data Reduction}
\newline
\justifying\let\raggedright\justifying

\normalsize
\textit{3.2.1. White Balance Calibration}
\newline

Since amateur astronomers are likely to exploit variable white balance setting in their cameras, the white balance of all the images and flat fields should be unified in the beginning. In practice, images in CR2/ RAW format are hard to adjust, while the use of other adjustable formats may result in a loss of information. Thus, the color temperature of the images were calibrated by the following algorithm.$$g_R=\frac{\overline{R_{2}}}{\overline{R_{1}}},\ g_G=\frac{\overline{G_{2}}}{\overline{G_{1}}},\ g_B=\frac{\overline{B_{2}}}{\overline{B_1}}$$We first outputted each raw image that was not adjusted as tif$_1$. Subsequently, the color temperature of the raw image was adjusted to $5770K$ and outputted as tif$_2$. The mean value of R channel reading of all the pixels in tif$_1$ was obtained as $\overline{R_1}$, and likewise for G and B channel, respectively. Next, by the mentioned formula, the mean value of R channel reading of all the pixels in the tif$_2$ was calculated as $\overline{R_2}$, and the gain on R channel as $g_R$; likewise, $g_G$ and $g_B$ were calculated. 
$$R_w=R_{raw}\cdot g_R,\ G_w=G_{raw}\cdot g_G,\ B_w=B_{raw}\cdot g_B$$Finally, with the mentioned formula, we obtained the white-balance-calibrated R channel reading of each pixel in the raw image $R_w$ , and likewise for $G_w$ and $B_w$ (Fig. 1). This specific color temperature was used as the most widely acknowledged ``normal’’ white is defined by the spectrum of an approximately 5770K black body (Philips 1995). With this algorithm, both the images of stars as labelled data and flat fields were calibrated. 
\begin{figure}[H]
\centering
\begin{minipage}[t]{0.37\linewidth}
\centering
\includegraphics[width=1.2\textwidth]{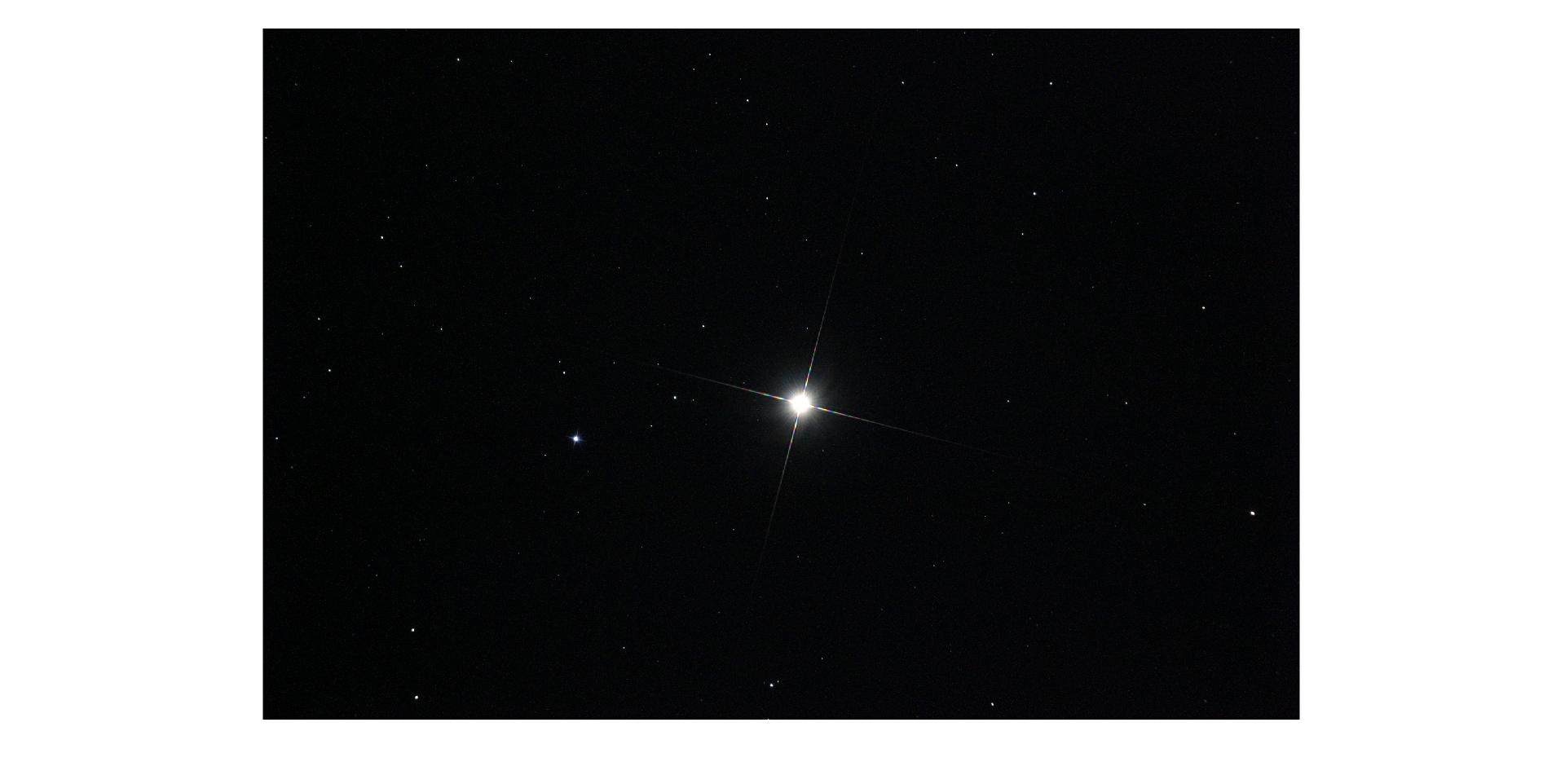}
\end{minipage}
\begin{minipage}[t]{0.37\linewidth}
\centering
\includegraphics[width=1.2\textwidth]{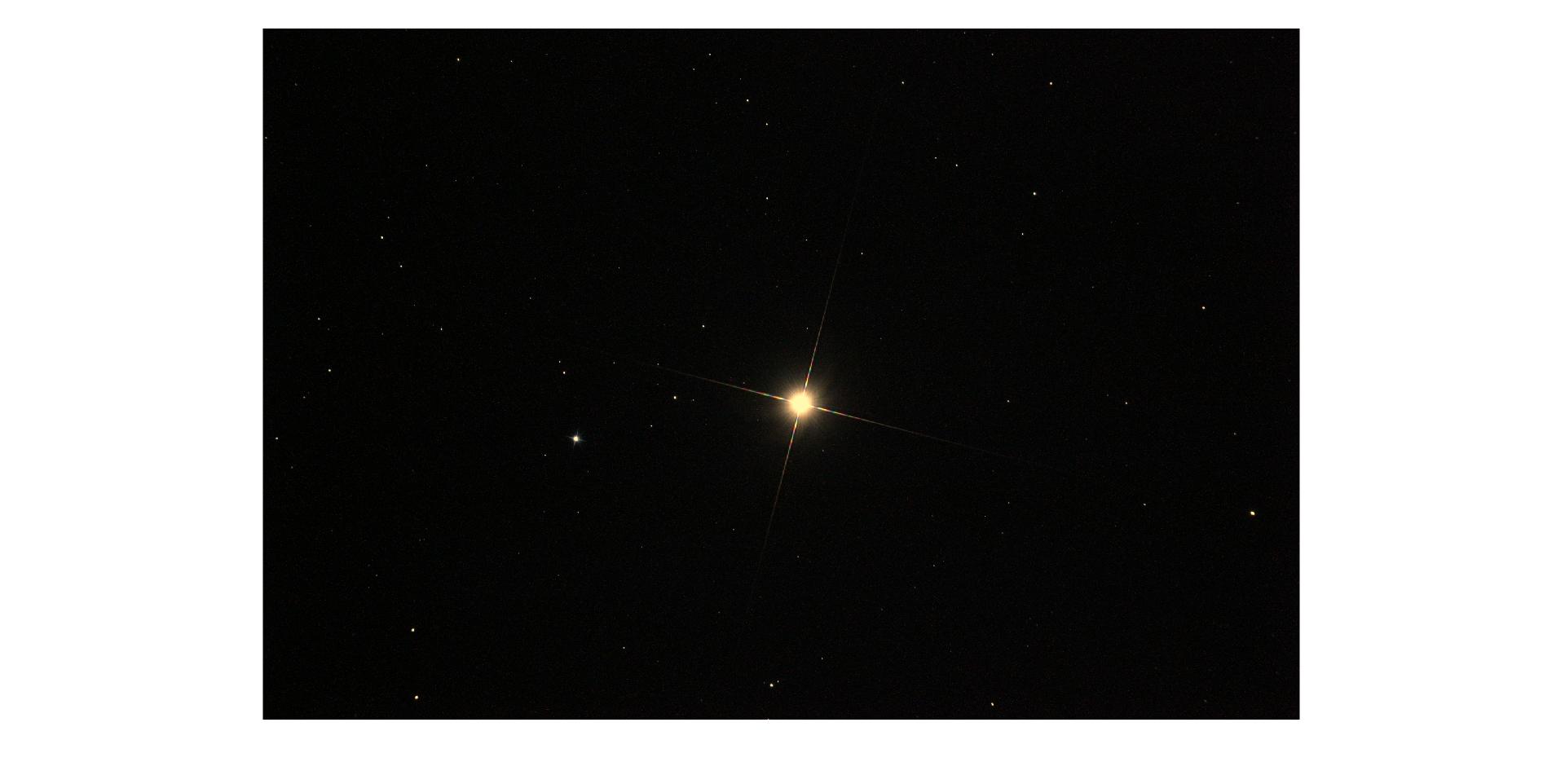}
\end{minipage}
\caption{\small{un-calibrated raw image (left) and white-balance calibrated image of HIP69672. Though this image did not finally act as labelled data for fewer stars in it, with its relatively low original color temperature, the effect on the white balance calibration could be more evidently shown.}}
\label{img}
\end{figure}

\textit{3.2.2. Flat Field Correction}
\newline

After white balance was calibrated for both stars and flat fields images, all the flat fields are averaged. Then, the common procedures was adopted to use flat field to eliminate the spatial non-uniformity of the CMOS caused by the entire imaging system. The R,G,B channel reading of each pixel in the image of stars after correcting the flat field are denoted as $R_f,\ G_f,\ B_f$, and R,G,B reading of the corresponding pixels in the averaged flat field are denoted as $R_{ff},\ G_{ff},\ B_{ff}$.$$R_f=\frac{R_w \overline{R_{ff}}}{R_{ff}},\ G_{f}=\frac{G_{w} \overline{G_{ff}}}{G_{ff}},\ B_{f}=\frac{B_{w} \overline{B_{ff}}}{B_{ff}}$$where $R_w,\ G_w,\ B_w$ denote the R,G,B channel reading of pixels in the white-balance-calibrated images, 
and $\overline{R_{ff}},\ \overline{G_{ff}},\ \overline{B_{ff}}$ indicate the mean value in R,G,B channels of all the pixels in the averaged flat field. 
\newline

\textit{3.2.3. Noise Cancellation}
\begin{figure}[H]
    \centering
    \includegraphics[width=.5\textwidth]{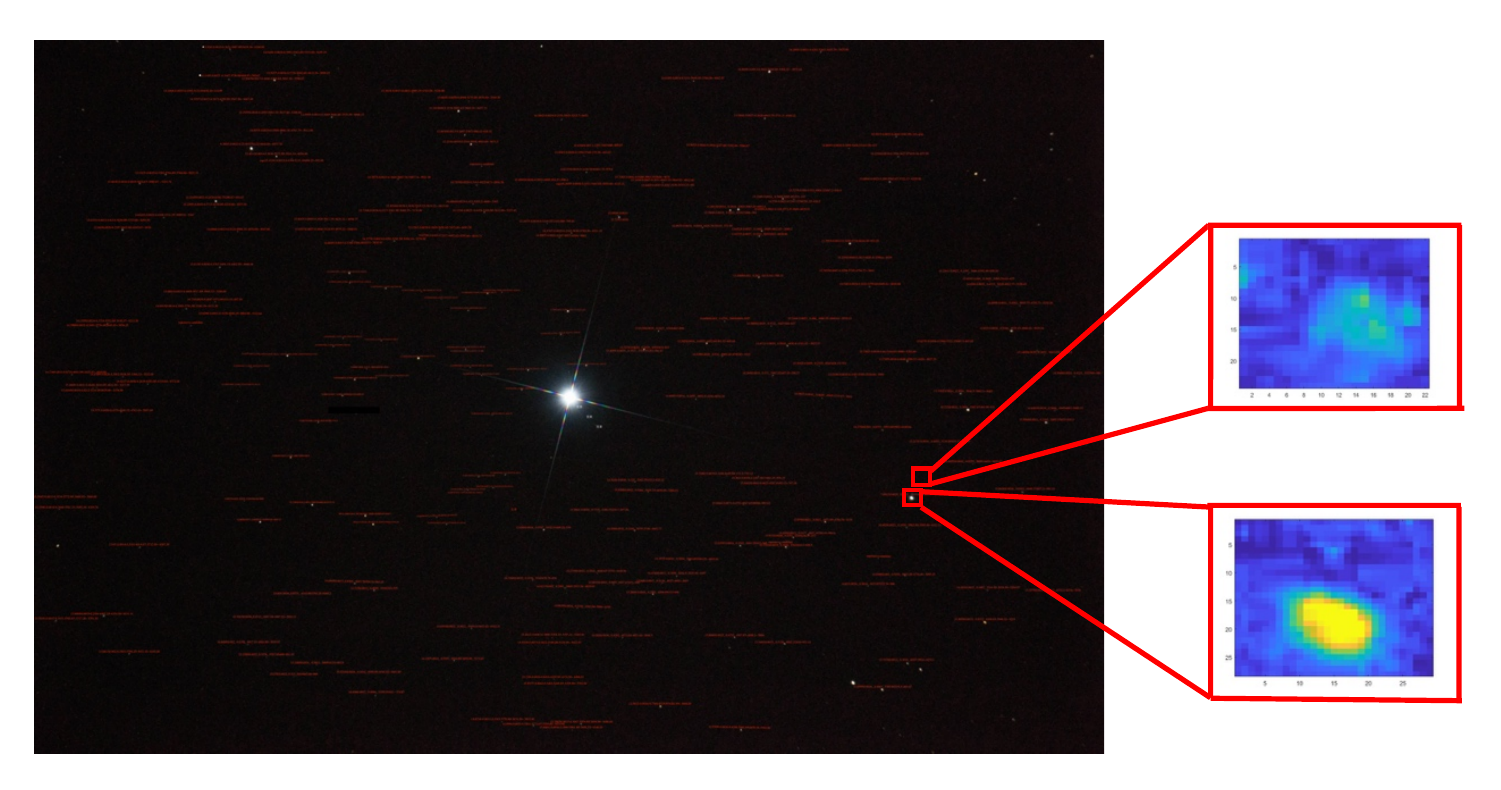}
    \caption{\small{After correcting the flat field, we cropped the image into “star patches” each sizing nearly 20$\times$20 pixels and containing a star respectively; a similar size of “background patch” was selected from the region surrounding each star patch correspondingly. A star patch and its background patch are presented in the lower red box and the upper red box, respectively.}}
    \label{img}
\end{figure}

After correcting the flat field and cropping the image into patches (Fig. 2), to enhance the reliability of labelled data by up-regulating their signal to noise ratio (SNR), the mean noise was subtracted from the star patches. The mean noise of each star patch in each channel was calculated with its corresponding background patch as the mean value of reading in R,G,B channel of all the pixels in the background patch $\overline{N_R},\ \overline{N_G},\ \overline{N_B}$. Subsequently, the mean noise was subtracted,$$R_m=R_f-\overline{N_R},\ G_m=G_f-\overline{N_G},\ B_m=B_f-\overline{N_B}$$The pixels with negative readings in any of the three channels that appeared after the mean was subtracted were considered noises and then set to 0, while the positive ones were left intact. Next, according to multiple trials, the pixels were set to 0 in the star patch with readings in any of the three channels that were smaller than the upper quartile (75\%) of those in the corresponding channel of the background patch. The choice of 75\% preserved sufficient information of the stars and effectively canceled most noises in the star patches. 
\newline

After the mentioned cancellation process, it was noticed that considerable bright heat noises appeared frequently at high ISO, which were still ``isolated pixels’’ in the star patches (Fig. 3). Another algorithm was adopted to set all pixels with all of adjacent pixels on their left, right, above, and below having 0 readings to 0, respectively. Thus, the effect of these bright heat noise pixels was reduced.
\newline

After all the noise cancellation, based on the observed results, we still discarded 29 of the faint stars with R,G,B reading comparable to noises in image that centers HIP65474. Subsequently, we discarded stars fainter than 15.5$^{th}$ mag. and some star patches with the sum of RGB readings over 50000, while used the rest 182 star patches for subsequent measurement. We aimed to remove all stars over-exposed (with either R,G,B reading over 255) by setting the upper threshold of 50000. The same procedure was implemented with the image centering HIP67301, and 145 star patches were kept. Finally, the reading of the pixels in the star patches was then summed channel by channel. 
\begin{figure}[H]
\centering
    \includegraphics[width=0.485\textwidth]{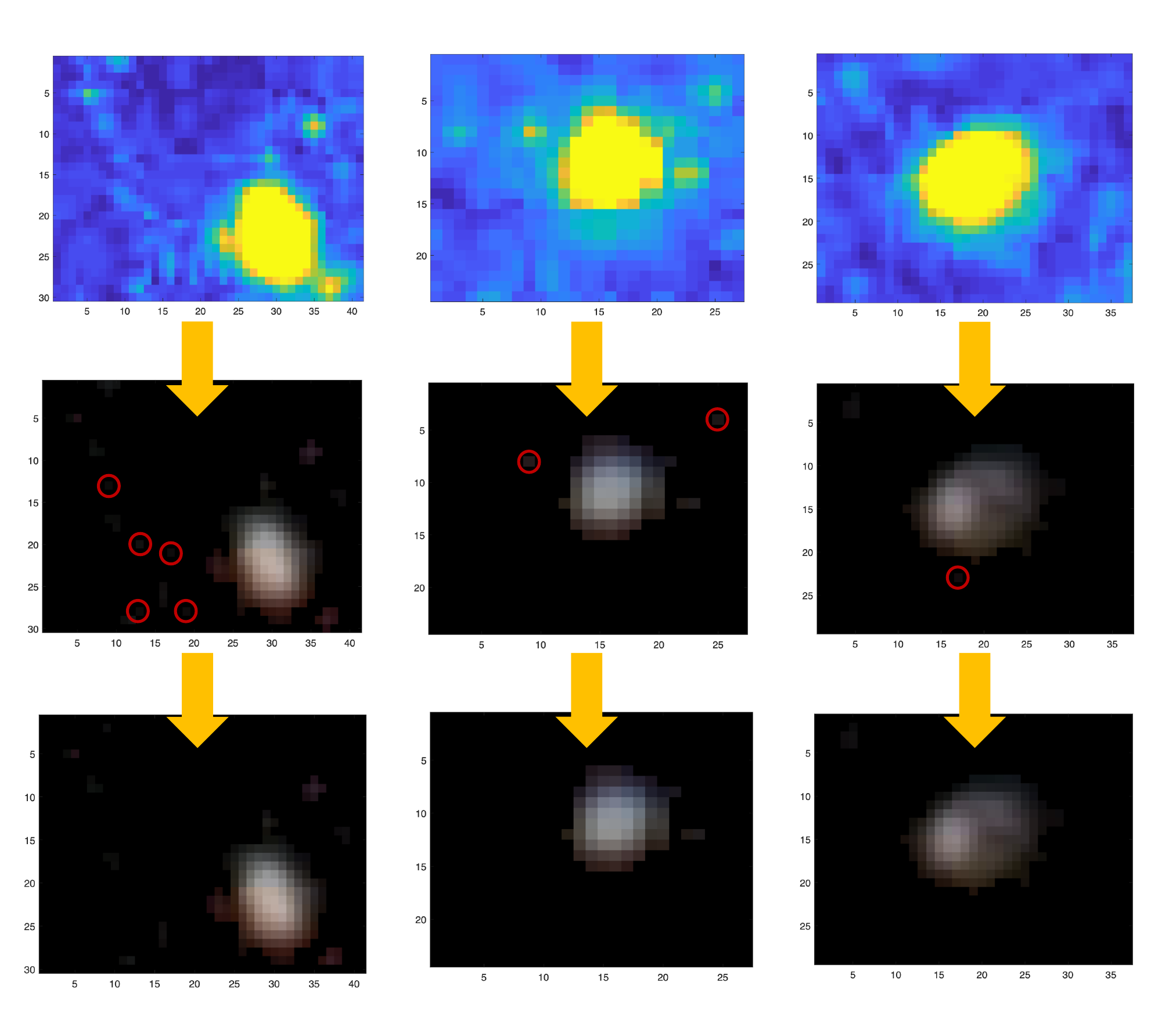}
    \caption{\small{3 star patches before noise cancellation, after mean noise and 75\%th percentile noise were subtracted, and after “isolated pixels” were removed. The red circles marked the presence of ``isolated pixels’’} }
    \label{img}
\end{figure}

\large
\textbf{4. Measurement for Apparent Magnitude with Function Fitting}
\newline
 
\normalsize
The apparent magnitude can be defined briefly as that the difference in magnitude of 5 indicates a 100 times proportion in intensities (greater value of magnitude corresponds to smaller intensity) (Carroll \& Ostile 2014),$$\frac{I_1}{I_2}=100^{(-\frac{m_1-m_2}{5})} \iff m_1-m_2=-2.5lg( \frac{I_1}{I_2}) $$where $I_1,\ I_2$ denote the intensities of two stars, respectively, and $m_1,\ m_2$ refer to their respective apparent magnitudes. As discussed above, two stars in one image exhibited roughly equal $\delta m$, so the mentioned equation still held with consideration of extinction. For CMOS, the $S = R + G + B$ of every star refers to a nonlinear function of intensity, so $m_1-m_2$ is a function about the natural logarithm of the ratio of $S$ of the two stars,$$m_1-m_2=f(-lg(\frac{S_1}{S_2}))=f(-ln(\frac{S_1}{S_2})/ln10)=kf(-ln(\frac{S_1}{S_2}))$$where $k$ is a constant, and $S_1,\ S_2$ denote the sum of R,G,B of the two stars, respectively. Given the relation above, the application of function fitting is considered to be feasible.
\newline

One bright star (apparent magnitude around 10$^{th}$ mag.) in one image that underwent the data reduction described above was taken as the reference star, and its RGB sum reading $S_0$ and expected output apparent magnitude $m_0$ were adopted for the reference. Based on the mentioned discussion, the measured magnitude $\hat{m}_i$ of a star with reading $S_i$ in the image can be yielded, $$\hat{m}_i=f(dm_i)+m_0,\ dm_i=-ln(\frac{S_i}{S_0})$$A strong logarithmic pattern was found according to the observed distribution of $\Delta m_i=m_i - m_0$ vs $dm_i=−ln(\frac{S_i}{S_0})$ where $S_i=R_i+G_i+B_i$ and $m_i$ is its corresponding expected output, both from each $i$th among the 181 ``star patches'' in the image centering HIP65474. Accordingly, the relation between $\Delta m_i$ and $dm_i$ was directly fitted by logarithm function,$$\hat{m}_i=f(dm_i)+m_0=aln(b\cdot dm_i+1)+m_0$$where $a,\ b$ represent coefficients to be determined by the data. Using this method, the absolute error $a_i$ of our fitted function $f$ on any $i$th star patch was assessed,$$a_i=|\hat{m}_i-m_i|$$Several attempts were made, covering regularized RBF network and ensemble learning the result to fit the data, whereas a larger error was yielded.
\newline

\large{\textbf{5. Measurement of Effective Temperature with Regularized RBF Network and Ensemble Learning}}
\begin{figure}[H]
    \centering
    \includegraphics[width=0.45\textwidth]{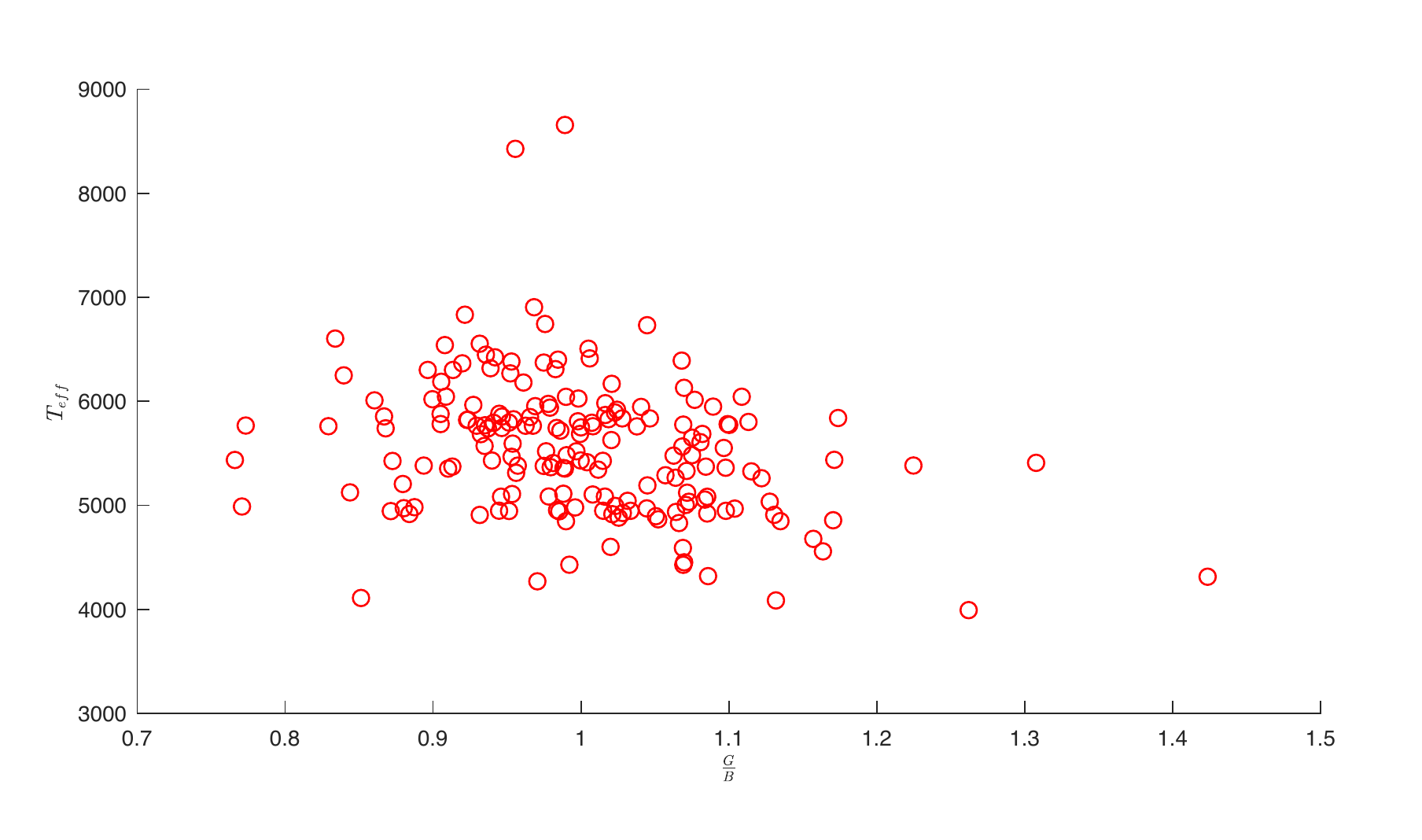}
    \caption{\small{The scatter plot of $T_{eff}$ vs $\frac{G}{B}$ showed no apparent pattern between the two variables, so the attempt to find the 1-dimensional relation between the two variables was discarded.}}
    \label{img}
\end{figure}

\normalsize
The theoretical relation between $T_{eff}$ and R, G, B readings is evident. Similar to UBV color system, R,G,B readings reveal the intensity of radiation from the star collected on the R,G,B wave bands. Since the effective temperature is critical to the stars’ black body spectra, it can be ascertained with R and B readings as a two-variable function,$$T_{eff}=F(R,B)$$The specific choice of using R and B instead of the ratio of $\frac{G}{B}$ or R, G, B was made by multiple attempts. We have considered that the ratio $\frac{G}{B}$ corresponds to the ratio of intensity collected in the B and V band in UBV color system and therefore corresponds to the color index $B – V$, which is commonly used for determining effective temperature; however, no obvious pattern was displayed in the distribution of $T_{eff}$ with $\frac{G}{B}$ of our data (Fig. 4). In brief, it was found that adopting R and B as a set of 2-dimensional inputs of the unknown function to measure the effective temperature yielded the highest accuracy. Given the availability of relatively sufficient labelled data, machine learning was used in this study as it can potentially give an accurate estimation of the unknown function. Specifically, we used regularized Radial Basis Function (RBF) network, i.e., a machine learning algorithm, to measure the effective temperature. 
\begin{figure}[H]
    \centering
    \includegraphics[width=.55\textwidth]{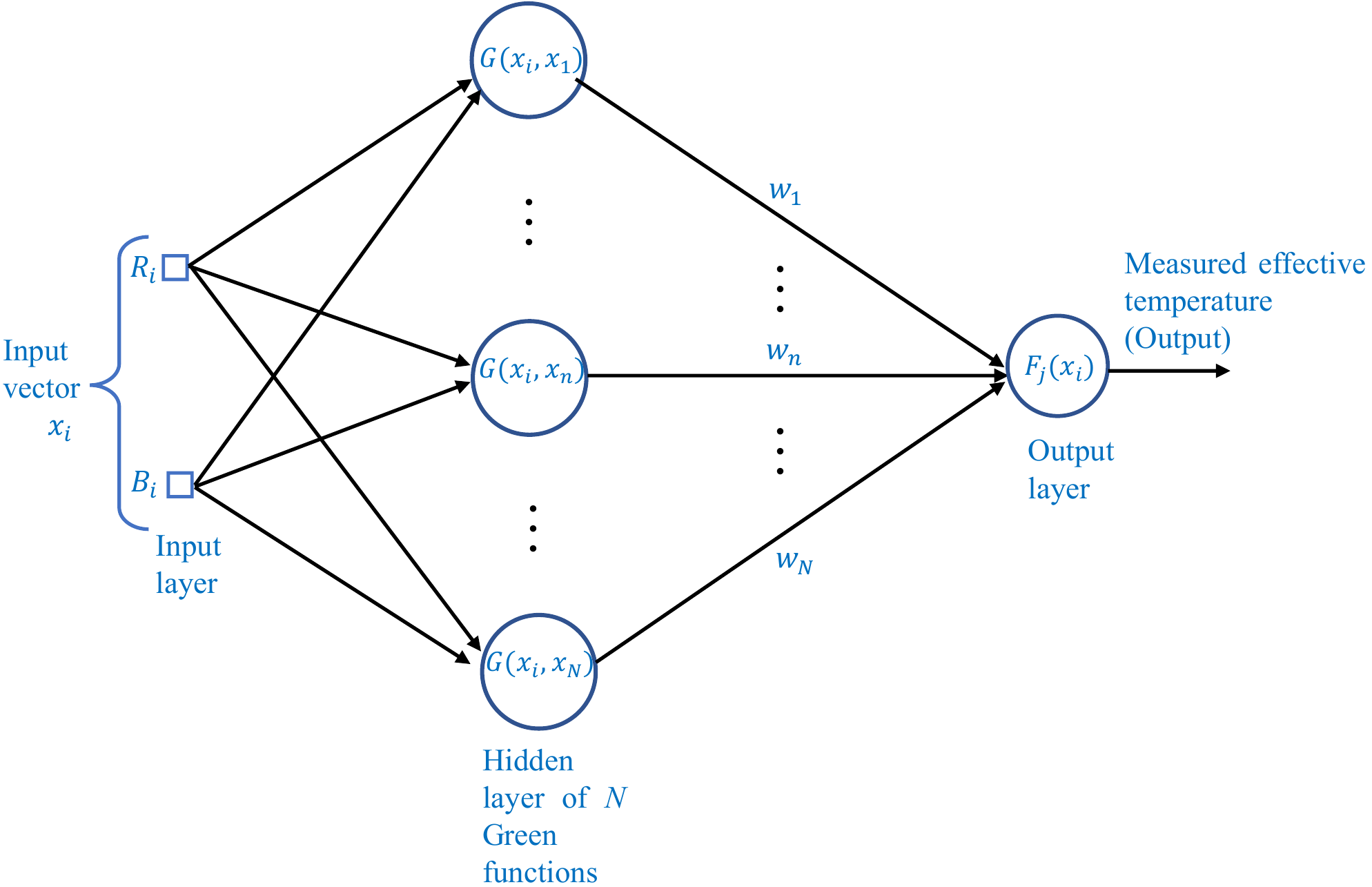}
    \caption{\small{The regularized RBF network. Each approximating function $F_j(x_i)=\sum\limits_{n=1}^{N} w_n G(x_i, x_n)$, where size of the training set $N$ equates number of labelled data. The expansion coefficients $w_n$ are determined based on the labelled data (Poggio \& Girosi 1990).}}
    \label{fig:my_label}
\end{figure}

We used a total of 145 labelled data from the image centering HIP67301 to train the RBF network to obtain $F(R, B)$. The labelled data were 2-dimensional vectors $x_i = (R_i,\ B_i),\ i = 1,\ 2,\ ...N$, and their corresponding expected output were $T_{eff_i}, i = 1,\ 2,\ ...,\ N$, where $N$ denotes the total number of labelled data (``star patches'') in the image. 80\% of the labelled data were randomly sampled to train the regularized RBF network, and one approximating function $F_j$ that measures the effective temperature was outputted (Fig. 5). The process was repeated for 1000 times, so 1000 different approximating functions $F_j,\ j = 1,\ 2,\ ...1000$ were yielded. With the $1000\ F_j$ , ensemble learning was conducted to output the measured effective temperature $\hat{T}_{eff}$ (Fig. 6). At a given unknown input RGB vector $x$, the measurement of the effective temperature could be calculated,$$\hat{T}_{eff}=\frac{\sum\limits_{j=1}^{1000} F_j(x)}{1000}$$
\begin{figure}[H]
  \centering
  \includegraphics[width=.4\textwidth]{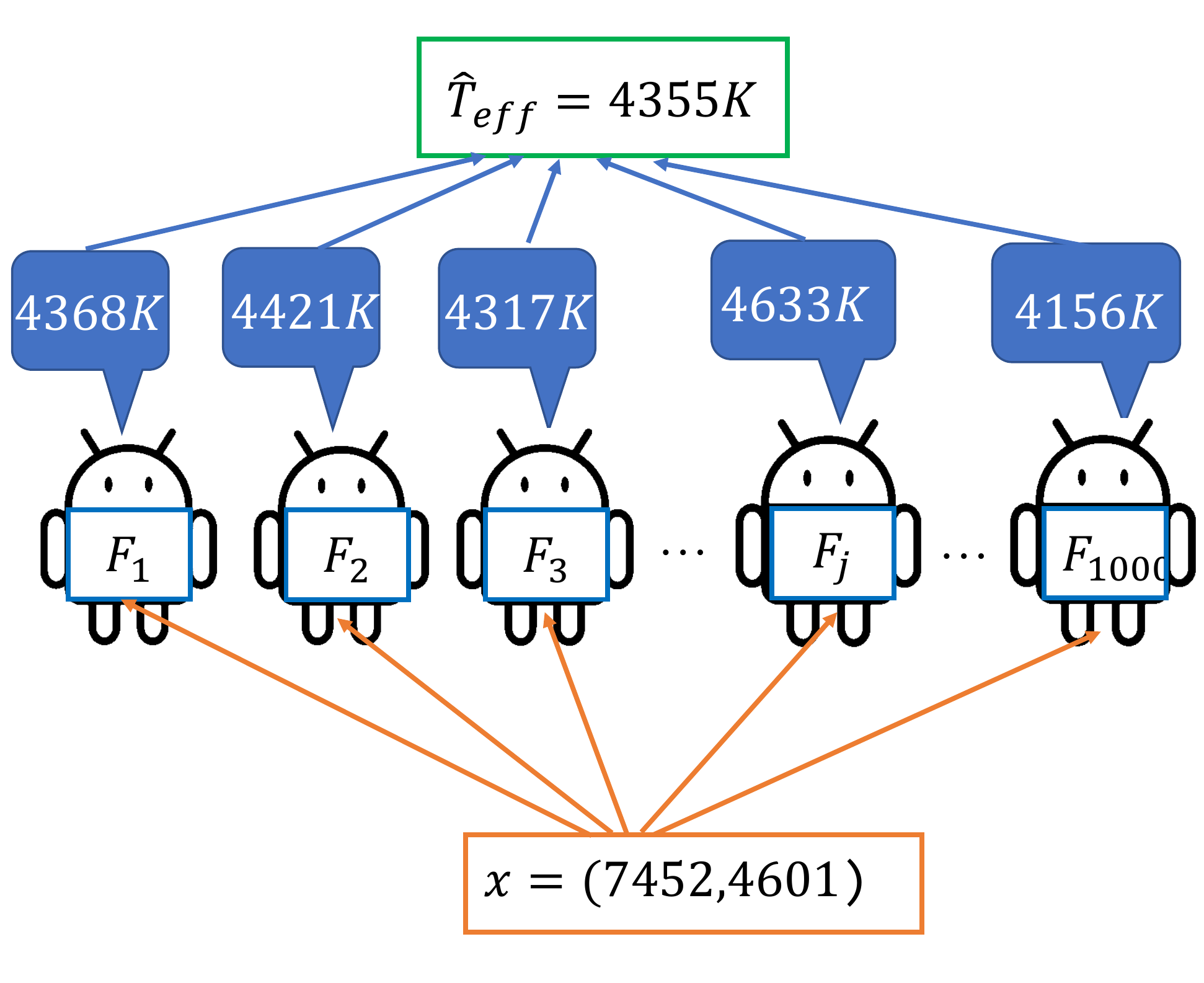} 
  \caption{\small{Measurement of the effective temperature with ensemble learning. The loss function of the regularized RBF network is defined as $L(F_j)=L_s(F_j)+\lambda L_c(F_j)$, where the regularization parameter $\lambda = 1\times 10^{-2}$, as calculated with the labelled data (Craven \& Wahba 1979); $L_s(F_j)=\frac{1}{2} \sum\limits_{i=1}^{N}(T_{eff_i}-F_j(x_i)^2)$ where $L_s(F_j)$ indicates the standard loss function, $L_c(F_j)=\frac{1}{2}||DF_j||^2$ where $L_c(F_j)$ refers to the regularization term (Tikhonov 1963).}}
  \label{img} 
\end{figure}

The generalization error of $\hat{T}_{eff}$ was assessed as below: for a specific labelled data $x_i$ and its corresponding expected output $T_{eff_i}$, there exist $k$ functions among the 1000 $F_j$s, with $x_i$ not included in their training sets, expressed as $F_{i1}, F_{i2}, ..., F_{ik}$. We are able to calculate the generalization error of $\hat{T}_{eff}$ tested by respective $x_i$ (Fig. 7),$$e_i=\frac{\sum\limits_{r=1}^k\frac{|F_{ir}(x_i)-{T_{eff}}_i|}{{T_{eff}}_i}}{k}$$Then, the mean generalization error of $\hat{T}_{eff}$ on all the labelled data is calculated,$$E_{\hat{T}_{eff}}=\frac{\sum\limits_{i=1}^{N} e_i}{N}$$In such a way, we have made use of every labelled data as testing data for once, hence improving the reliability of the generalization error and achieved higher utility of our limited data. 
\begin{figure}[H]
\centering
\includegraphics[width=.4 \textwidth]{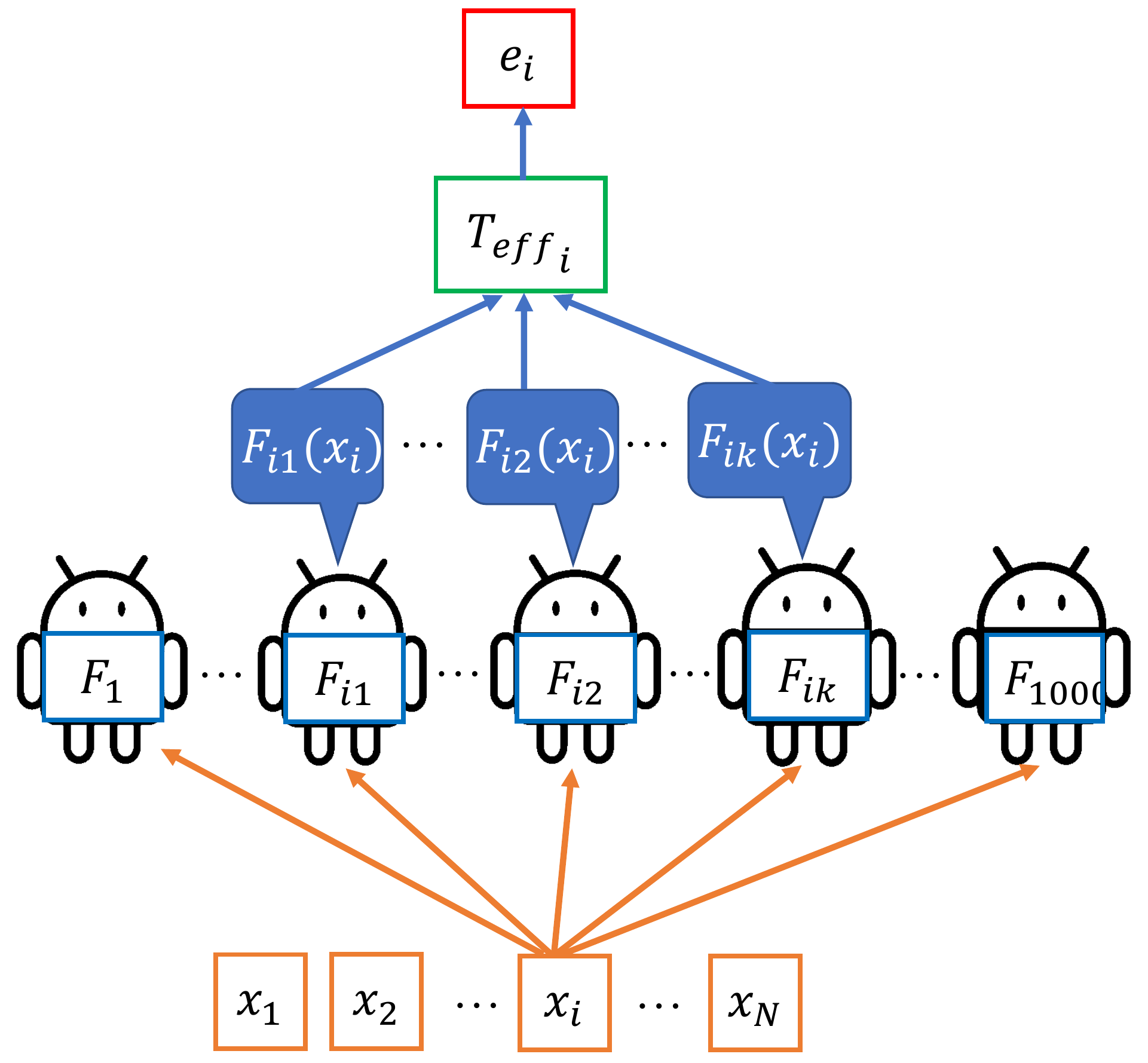}
\caption{\small{Calculation of generalization error $e_i$.}}
\label{img}
\end{figure}

\large{\textbf{6. Results}}
\newline

\textit{6.1. Magnitude Measurement}
\newline

\normalsize
Data from the 181 star patches in HIP65474 were adopted to perform the function fitting algorithm, and the function below was yielded to measure the apparent magnitude (Fig. 8),$$\hat{m}_i=3.0907ln(0.7970\cdot ln(\frac{S_i}{S_0})+1)+m_0$$Due to different atmospheric extinction effect on stars at various altitudes, we used data from HIP65474 solely.
\begin{figure}[H]
\centering
\includegraphics[width=0.6\textwidth]{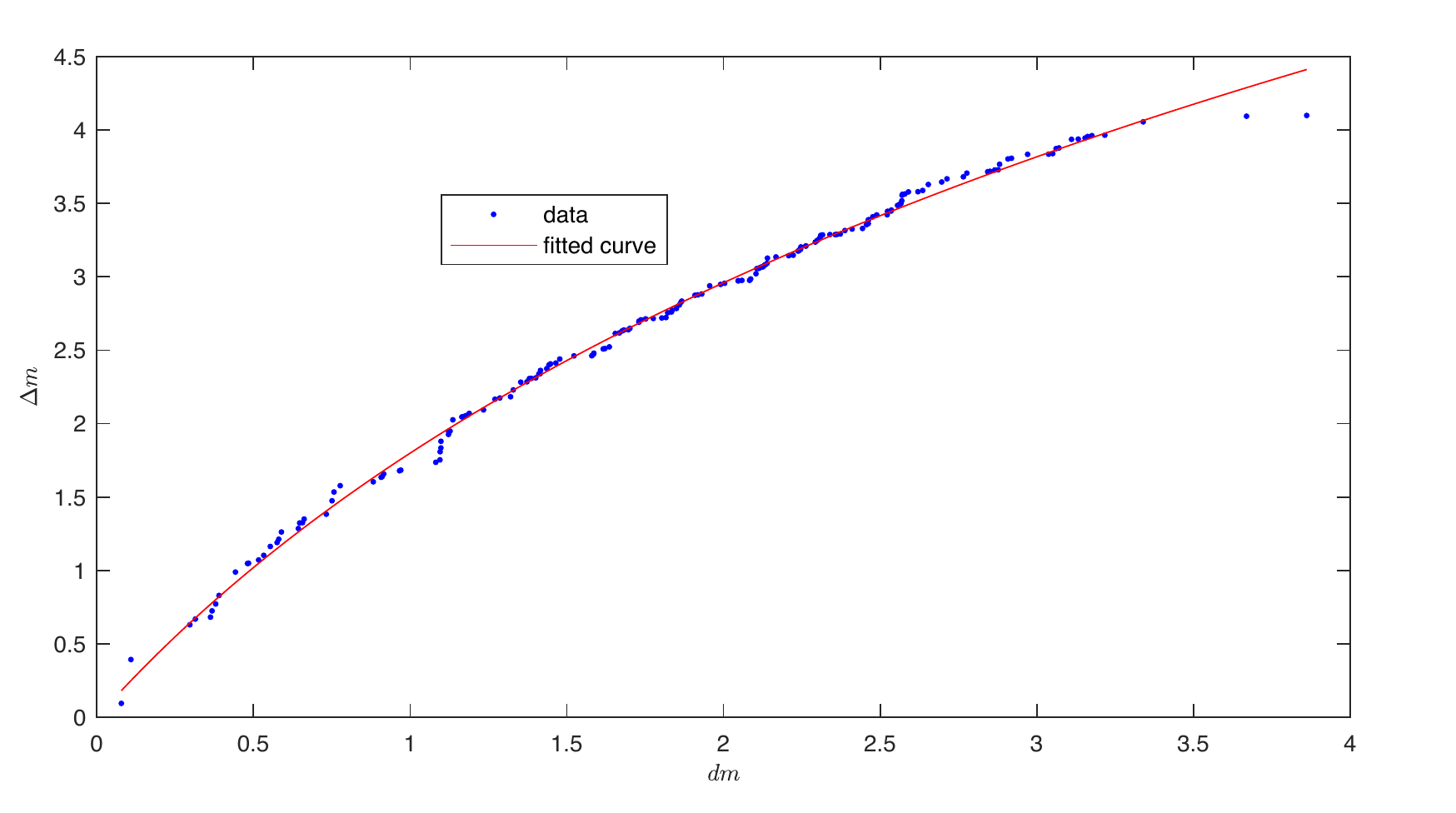}
\caption{\small{The original distribution of $\Delta m$ vs $dm$ and the logarithmic function fitted $\hat{m} – m_0 = f (dm)$ using the data from HIP65474.}}
\label{img}
\end{figure}
\leavevmode
\newline

To display the distribution of the absolute error of the function in measuring magnitude, we plotted the histogram with 10 classes (Fig. 9). The mean of all the absolute errors $a_i$ was calculated, which is 0.1035$^{th}$mag., and its 90\% confidence interval, which is $0.1035\pm 0.0106^{th}$mag.. 
\begin{figure}[H]
\centering
\includegraphics[width=0.55\textwidth]{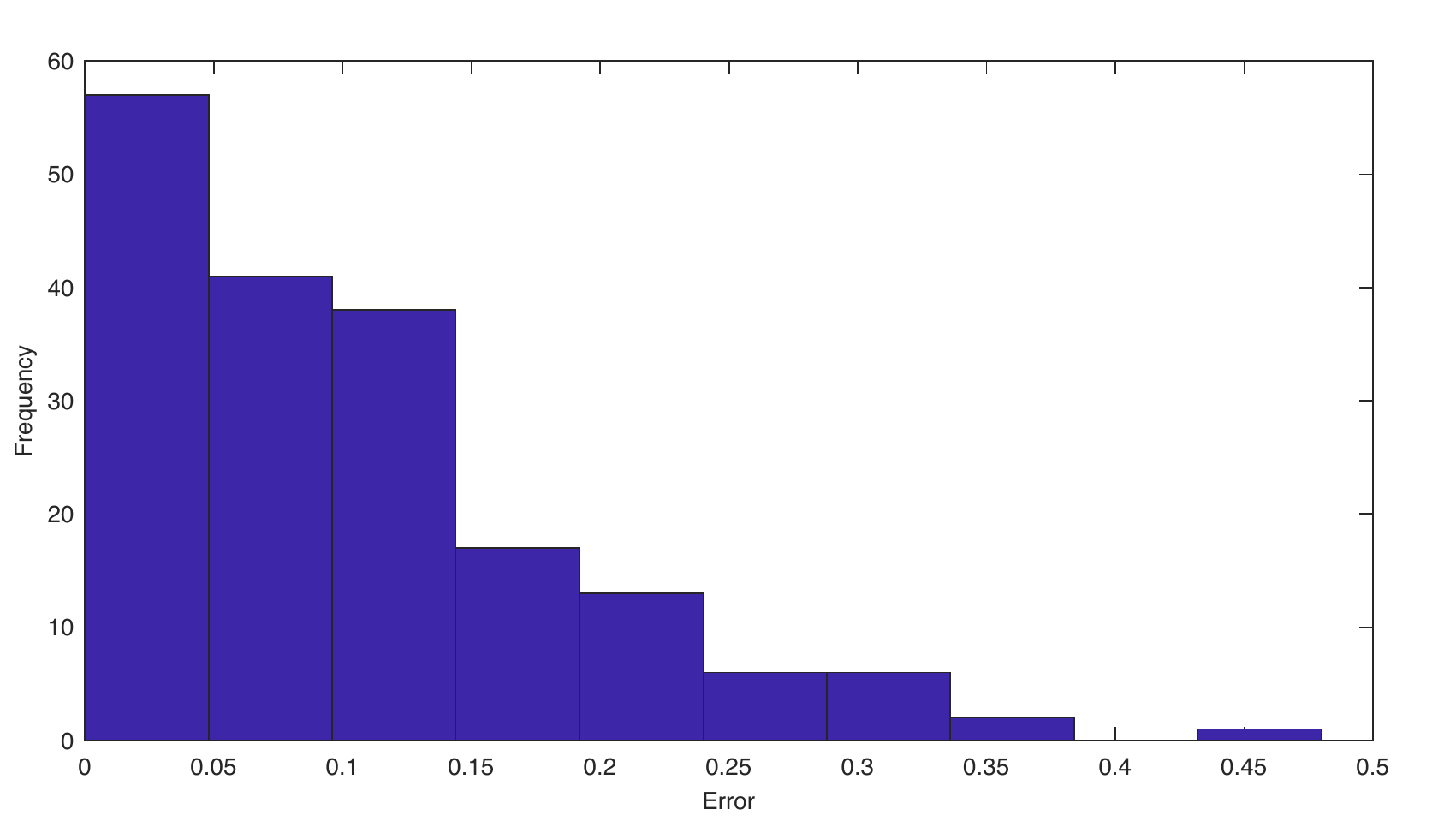}
\caption{\small{The histogram of $\hat{m}$ with absolute error $a_i$ with 10 equal classes of the 181 labelled from HIP 65474.}}
\label{img}
\end{figure}

\large
\textit{6.2. Effective Temperature Measurement}
\newline

\normalsize
To measure the effective temperature, we employed labelled data from 145 ``star patches’’ in the image centering HIP67301. The generalization errors in the measured $\hat{T}_{eff}$ concentrated on relatively smaller values (Fig. 10). Furthermore, despite some individual measured effective temperature with abnormally large errors, the mean generalization error $E_{\hat{T}_{eff}}$ exhibits 0.07 and 90\% confidence interval of $e_i$ is $0.07\pm0.01$.
\begin{figure}[H]
  \centering
\includegraphics[width=0.55\textwidth]{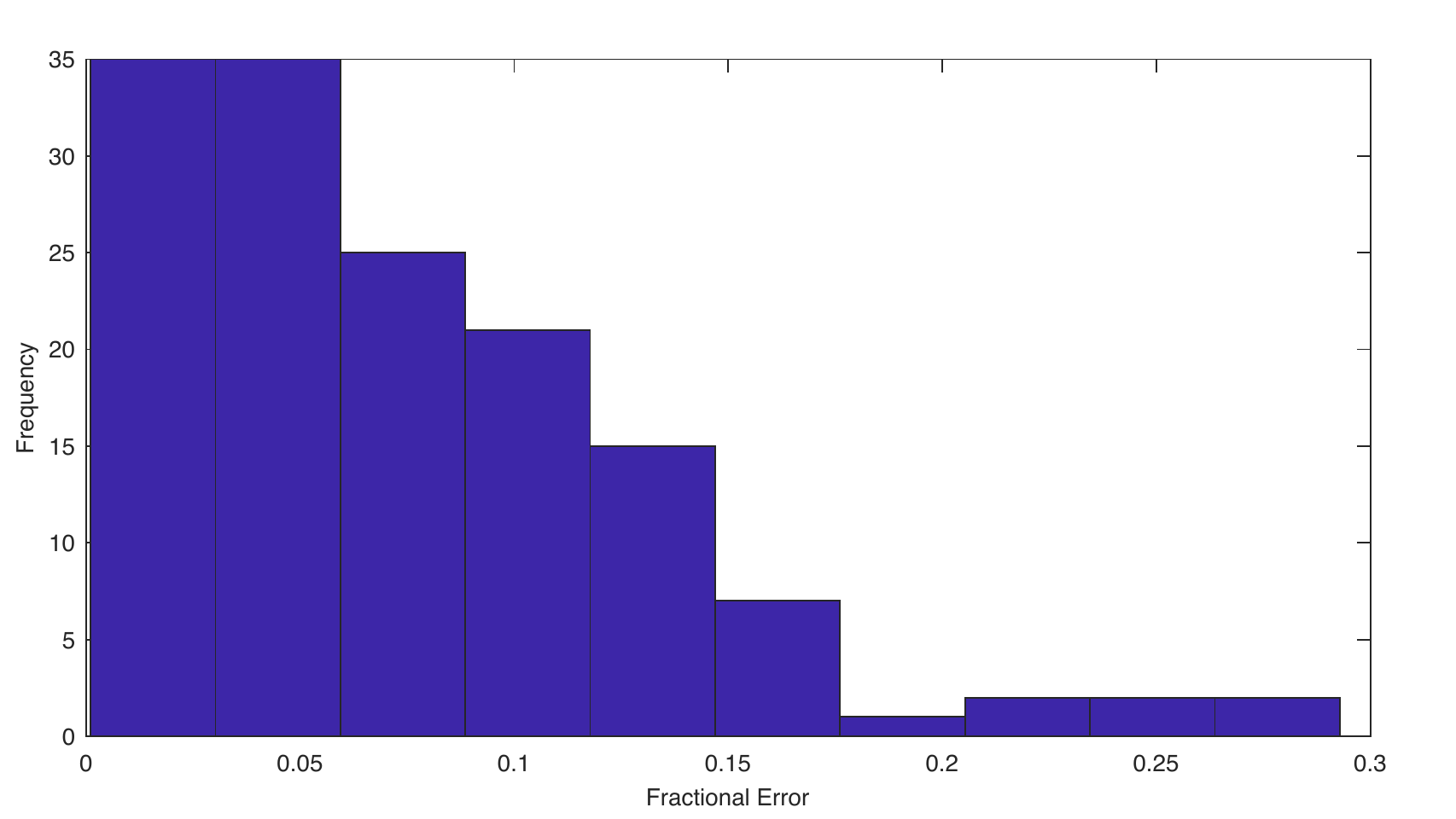}
\caption{\small{The histogram of generalization error $e_i$ with 10 equal classes of the 145 testing data from HIP 67301.}}
\label{img}
\end{figure}

As $F(x),\ x = (R, B)$ depends on the relation between R and B, $\hat{T}_{eff}$ trained with one image can be adopted to measure with unknown data from another image. For verification, we measured the effective temperature with $x_i$ from image centering HIP65474 using $\hat{T}_{eff}$ function trained with data from HIP67301, and we yielded the mean generalization error of 0.09 and 90\% confidence interval of $0.09\pm0.01$ (Fig. 11). 
\begin{figure}[H]
    \centering
    \includegraphics[width=0.55\textwidth]{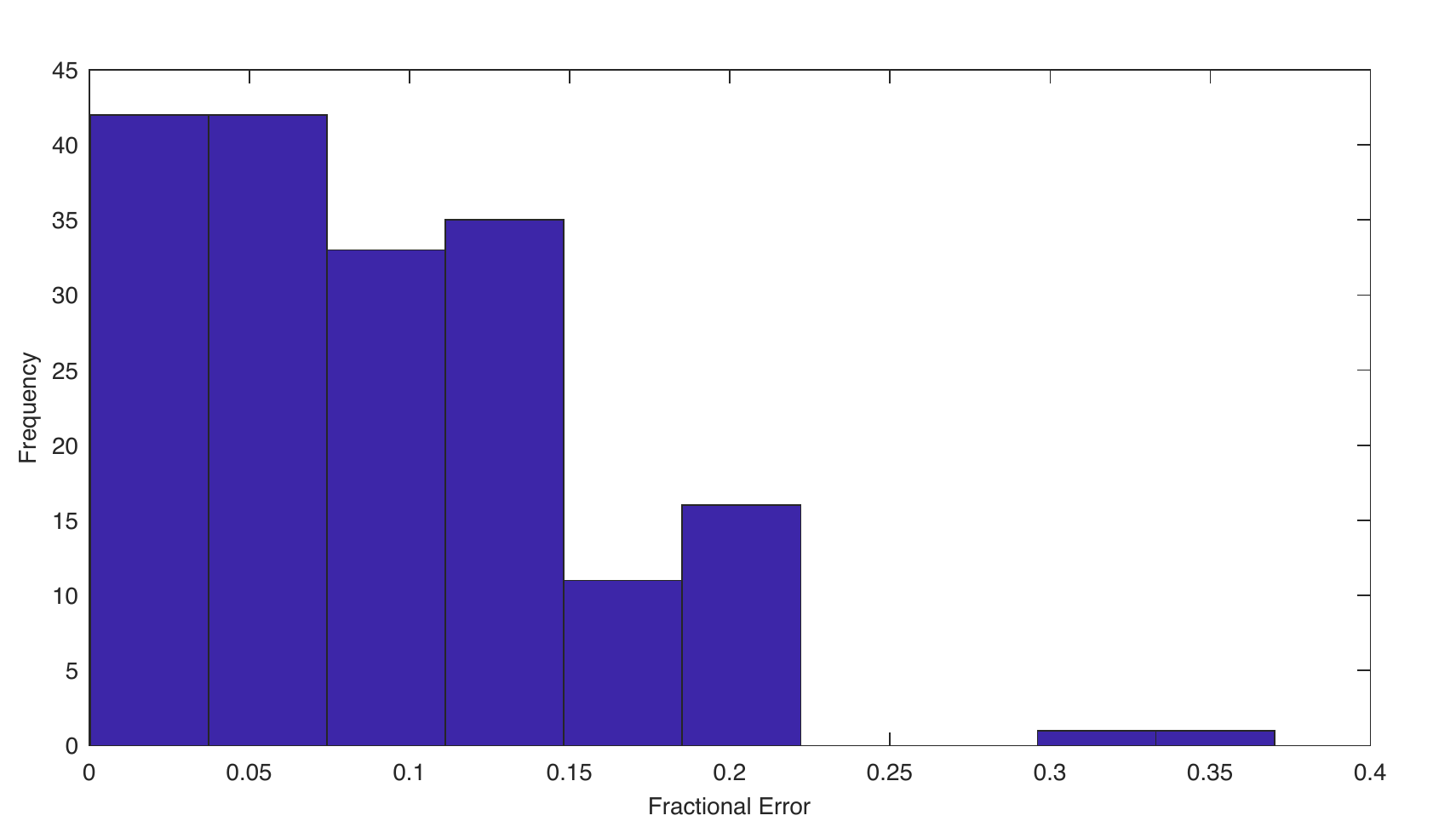}
    \caption{\small{The histogram of the fractional errors of $\hat{T}_{eff}$ with 10 classes. The function trained with data from HIP67301 are used to measure the effective temperature of the 181 stars in the image of HIP65474 with their R and B data.}}
    \label{fig:my_label}
\end{figure}
\large{\textbf{7. Conclusions and Discussion}}
\newline

\normalsize
As discussed above, one algorithm was developed to measure the apparent magnitudes, while the other was yielded to measure the effective temperature with data from amateur telescopes. Our measurement of magnitudes exhibited absolute error at about 0.1 and measurement of the effective temperature displayed a mean generalization error at nearly 0.09. The algorithm for measuring effective temperature can be used on different images shot at the same geographic location and within a short period of time. Both algorithms are compatible with various white balances and do not require the use of dark fields. 
\newline

It was reported that there exist a ``v’’ shape distribution of the generalization errors and corresponding effective temperature (Fig.12), indicating $\hat{T}_{eff}$  is more inaccurate when the star has relatively lower ($< 4750K$) or higher ($> 6000K$) $T_{eff}$. 
\begin{figure}[H]
\centering
\includegraphics[width=0.55\textwidth]{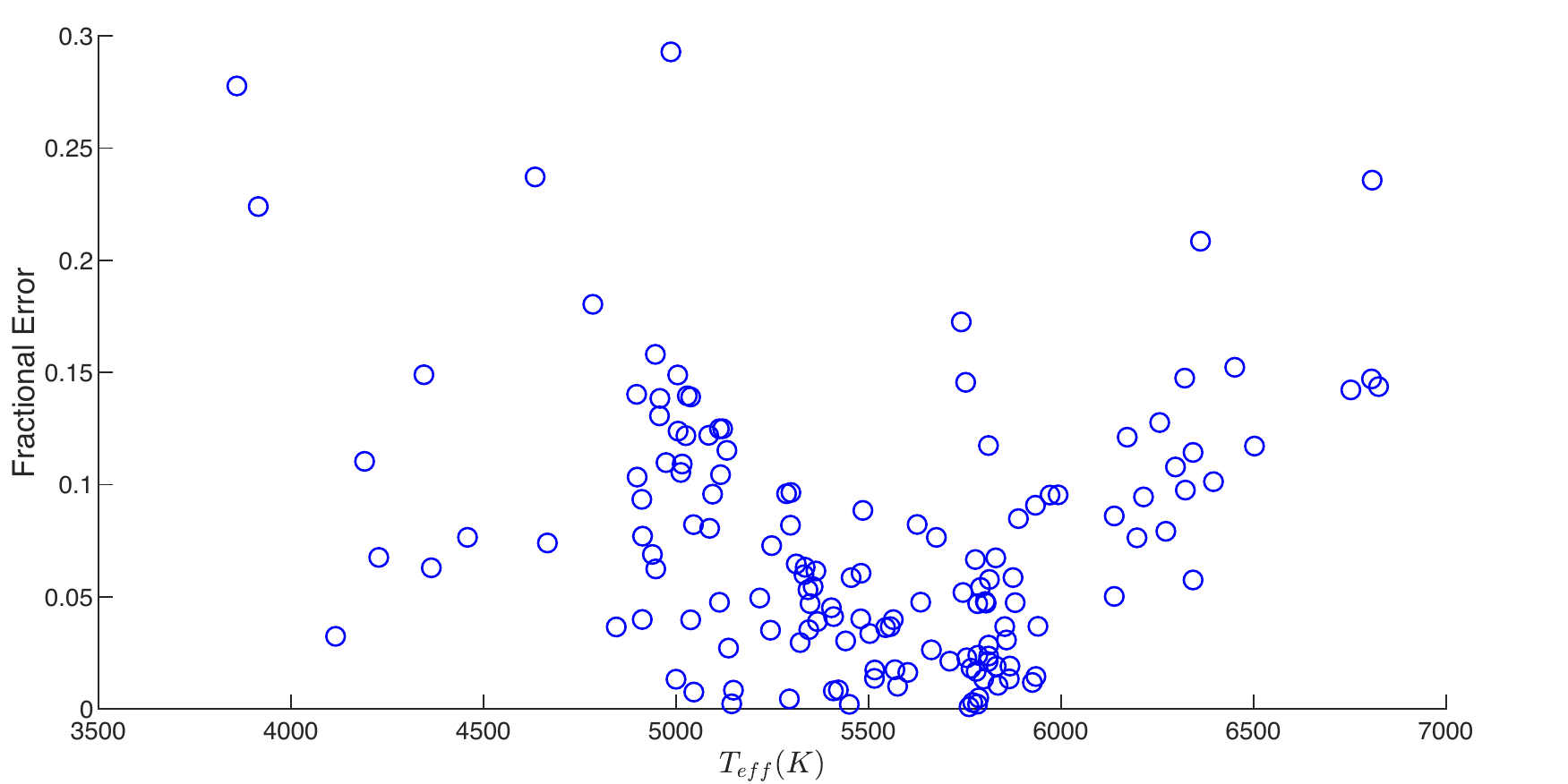}
\caption{\small{The distribution of generalization error vs expected output $T_{eff}$ with data from HIP67301.}}
\end{figure}

We hypothesized that this ``v’’ shape distribution of generalization error is caused by relative scarcity of labelled data with $T_{eff}< 4750K$ and $T_{eff}> 6000K$, which is supported by the distribution of labelled data (Fig. 13). As a result, the accuracy of $\hat{T}_{eff}$ can be enhanced when more labelled data that has $T_{eff}$ out of the interval [$4750K,\ 6000K$] are incorporated into future researches. 
\begin{figure}[H]
\centering
\includegraphics[width=0.55\textwidth]{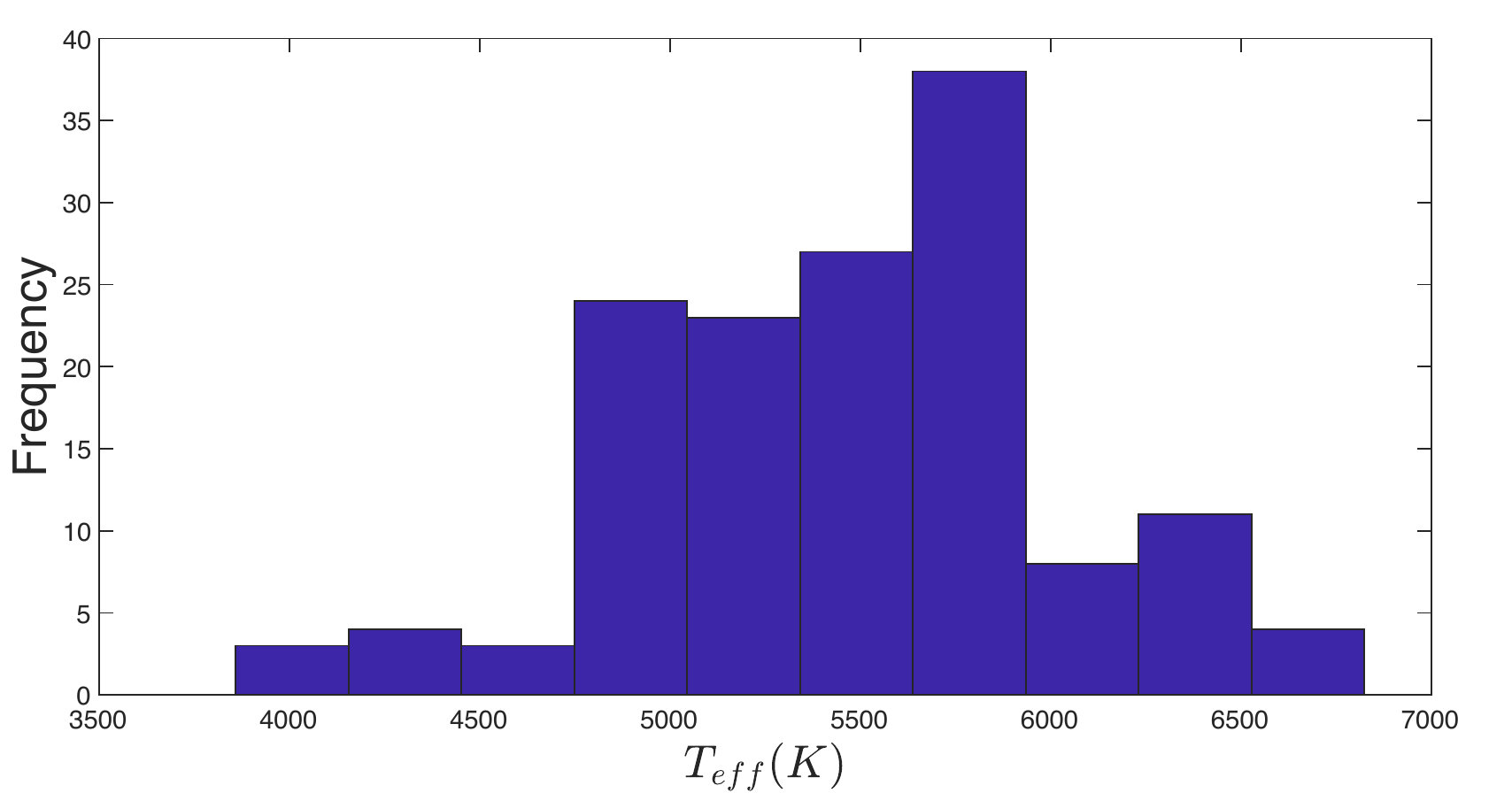}
\caption{\small{The histogram of expected output $T_{eff}$ with 10 classes in the image centering HIP67301.}}
\label{img}
\end{figure}

\large{\textbf{Appendix}}
\newline
\scriptsize
\begin{longtable}{rrrrr}
  \caption{\small{Measurement of apparent magnitude with labelled data from the image centering HIP65474.}}
  \\ 
  \hline
  \multicolumn{1}{l}{$S_i$} & \multicolumn{1}{l}{$m_i$} & \multicolumn{1}{l}{$\hat{m}$} & \multicolumn{1}{l}{$a_i$}\\
    \hline
    \endhead
    25848 & 12.7519  & 12.6529  & 0.0990  \\
    16072 & 13.3279  & 13.3684  & 0.0405  \\
    5140  & 14.4830  & 14.6130  & 0.1300  \\
    3539  & 14.8210  & 14.9314  & 0.1104  \\
    9621  & 14.0388  & 13.9911  & 0.0477  \\
    5243  & 14.4920  & 14.5951  & 0.1031  \\
    11667 & 13.8626  & 13.7715  & 0.0911  \\
    7163  & 14.2754  & 14.2994  & 0.0240  \\
    4100  & 14.6483  & 14.8098  & 0.1615  \\
    18793 & 13.1382  & 13.1507  & 0.0125  \\
    13186 & 13.7395  & 13.6236  & 0.1159  \\
    4151  & 14.6898  & 14.7993  & 0.1095  \\
    23311 & 12.9359  & 12.8231  & 0.1128  \\
    29468 & 12.5652  & 12.4224  & 0.1428  \\
    5968  & 14.5850  & 14.4758  & 0.1092  \\
    11270 & 13.8006  & 13.8121  & 0.0115  \\
    16519 & 13.3493  & 13.3313  & 0.0180  \\
    16466 & 13.2809  & 13.3357  & 0.0548  \\
    2855  & 15.0680  & 15.1008  & 0.0328  \\
    5867  & 14.3746  & 14.4918  & 0.1172  \\
    7321  & 14.3572  & 14.2776  & 0.0796  \\
    16741 & 13.2359  & 13.3131  & 0.0772  \\
    25483 & 12.7253  & 12.6769  & 0.0484  \\
    12048 & 13.6310  & 13.7334  & 0.1024  \\
    6302  & 14.4590  & 14.4242  & 0.0348  \\
    22710 & 12.9790  & 12.8647  & 0.1143  \\
    2769  & 15.0464  & 15.1242  & 0.0778  \\
    2062  & 15.3386  & 15.3410  & 0.0024  \\
    8569  & 14.1080  & 14.1158  & 0.0078  \\
    3806  & 14.9188  & 14.8719  & 0.0469  \\
    5220  & 14.7628  & 14.5991  & 0.1637  \\
    6658  & 14.4685  & 14.3713  & 0.0972  \\
    3792  & 14.9802  & 14.8750  & 0.1052  \\
    7767  & 14.2840  & 14.2178  & 0.0662  \\
    2876  & 15.3652  & 15.0952  & 0.2700  \\
    5225  & 14.5280  & 14.5983  & 0.0703  \\
    23167 & 12.7274  & 12.8331  & 0.1057  \\
    25609 & 12.5920  & 12.6686  & 0.0766  \\
    15044 & 13.4711  & 13.4560  & 0.0151  \\
    9756  & 13.9126  & 13.9757  & 0.0631  \\
    4924  & 14.5889  & 14.6514  & 0.0625  \\
    4957  & 14.6884  & 14.6454  & 0.0430  \\
    3704  & 15.1164  & 14.8943  & 0.2221  \\
    13860 & 13.4588  & 13.5613  & 0.1025  \\
    5268  & 14.5480  & 14.5908  & 0.0428  \\
    2154  & 15.2350  & 15.3098  & 0.0748  \\
    44291 & 11.7953  & 11.5713  & 0.2240  \\
    10770 & 13.8737  & 13.8645  & 0.0092  \\
    15264 & 13.4524  & 13.4370  & 0.0154  \\
    30429 & 12.3908  & 12.3632  & 0.0276  \\
    5350  & 14.6925  & 14.5768  & 0.1157  \\
    19767 & 13.1549  & 13.0769  & 0.0780  \\
    4908  & 14.5760  & 14.6543  & 0.0783  \\
    2090  & 15.3456  & 15.3314  & 0.0142  \\
    19864 & 13.0804  & 13.0697  & 0.0107  \\
    2532  & 14.9653  & 15.1916  & 0.2263  \\
    3783  & 14.8093  & 14.8769  & 0.0676  \\
    7654  & 14.2354  & 14.2327  & 0.0027  \\
    13646 & 13.5844  & 13.5809  & 0.0035  \\
    9054  & 14.1613  & 14.0571  & 0.1042  \\
    8755  & 14.0189  & 14.0930  & 0.0741  \\
    7694  & 14.2095  & 14.2274  & 0.0179  \\
    7635  & 14.0994  & 14.2353  & 0.1359  \\
    3331  & 14.8234  & 14.9802  & 0.1568  \\
    11609 & 13.7085  & 13.7774  & 0.0689  \\
    2201  & 15.2360  & 15.2943  & 0.0583  \\
    8695  & 14.1207  & 14.1003  & 0.0204  \\
    6746  & 14.6704  & 14.3585  & 0.3119  \\
    3745  & 14.9629  & 14.8852  & 0.0777  \\
    4098  & 14.8217  & 14.8102  & 0.0115  \\
    4210  & 14.7260  & 14.7874  & 0.0614  \\
    3835  & 14.6832  & 14.8656  & 0.1824  \\
    2666  & 15.2739  & 15.1530  & 0.1209  \\
    18718 & 13.0360  & 13.1564  & 0.1204  \\
    4207  & 14.9007  & 14.7880  & 0.1127  \\
    6373  & 14.3767  & 14.4135  & 0.0368  \\
    2368  & 15.1673  & 15.2411  & 0.0738  \\
    1039  & 15.4993  & 15.7930  & 0.2937  \\
    7987  & 14.0153  & 14.1891  & 0.1738  \\
    1979  & 15.4943  & 15.3700  & 0.1243  \\
    6984  & 14.3862  & 14.3245  & 0.0617  \\
    5809  & 14.4568  & 14.5011  & 0.0443  \\
    2312  & 15.2783  & 15.2586  & 0.0197  \\
    1259  & 15.3369  & 15.6728  & 0.3359  \\
    3593  & 15.1298  & 14.9191  & 0.2107  \\
    3964  & 14.8469  & 14.8381  & 0.0088  \\
    12765 & 13.7130  & 13.6635  & 0.0495  \\
    9811  & 14.0405  & 13.9695  & 0.0710  \\
    5427  & 14.6045  & 14.5638  & 0.0407  \\
    7873  & 14.2779  & 14.2039  & 0.0740  \\
    1751  & 15.4557  & 15.4550  & 0.0007  \\
    3911  & 15.1281  & 14.8493  & 0.2788  \\
    2337  & 15.2036  & 15.2508  & 0.0472  \\
    9183  & 14.1234  & 14.0418  & 0.0816  \\
    6010  & 14.6864  & 14.4692  & 0.2172  \\
    4288  & 14.7873  & 14.7718  & 0.0155  \\
    4986  & 14.7559  & 14.6402  & 0.1157  \\
    30532 & 12.2316  & 12.3569  & 0.1253  \\
    11999 & 13.8636  & 13.7382  & 0.1254  \\
    2696  & 15.3618  & 15.1445  & 0.2173  \\
    11413 & 13.8811  & 13.7974  & 0.0837  \\
    11985 & 13.7635  & 13.7396  & 0.0239  \\
    20439 & 13.0408  & 13.0271  & 0.0137  \\
    8752  & 13.9100  & 14.0934  & 0.1834  \\
    5893  & 14.3396  & 14.4877  & 0.1481  \\
    5849  & 14.3770  & 14.4947  & 0.1177  \\
    5252  & 14.4217  & 14.5936  & 0.1719  \\
    5928  & 14.4789  & 14.4821  & 0.0032  \\
    4872  & 14.7160  & 14.6608  & 0.0552  \\
    3774  & 14.7898  & 14.8789  & 0.0891  \\
    8138  & 14.1559  & 14.1698  & 0.0139  \\
    3074  & 15.0296  & 15.0436  & 0.0140  \\
    6025  & 14.0915  & 14.4669  & 0.3754  \\
    3915  & 14.7300  & 14.8485  & 0.1185  \\
    12167 & 13.8138  & 13.7216  & 0.0922  \\
    27781 & 12.6636  & 12.5282  & 0.1354  \\
    4652  & 14.6886  & 14.7014  & 0.0128  \\
    6128  & 14.6102  & 14.4508  & 0.1594  \\
    3476  & 15.2395  & 14.9460  & 0.2935  \\
    12338 & 13.6838  & 13.7048  & 0.0210  \\
    13072 & 13.7412  & 13.6344  & 0.1068  \\
    2324  & 15.4334  & 15.2549  & 0.1785  \\
    28987 & 12.4511  & 12.4523  & 0.0012  \\
    3780  & 14.8555  & 14.8776  & 0.0221  \\
    2108  & 15.3559  & 15.3253  & 0.0306  \\
    12515 & 13.4950  & 13.6876  & 0.1926  \\
    4679  & 14.5797  & 14.6963  & 0.1166  \\
    2291  & 15.1068  & 15.2653  & 0.1585  \\
    33436 & 12.5045  & 12.1827  & 0.3218  \\
    9235  & 14.1145  & 14.0357  & 0.0788  \\
    5137  & 14.5448  & 14.6136  & 0.0688  \\
    10108 & 13.9240  & 13.9363  & 0.0123  \\
    36056 & 12.0838  & 12.0301  & 0.0537  \\
    3959  & 14.8872  & 14.8392  & 0.0480  \\
    34186 & 12.1736  & 12.1386  & 0.0350  \\
    34358 & 12.0319  & 12.1285  & 0.0966  \\
    31728 & 12.1266  & 12.2844  & 0.1578  \\
    5975  & 14.4593  & 14.4747  & 0.0154  \\
    4536  & 14.6600  & 14.7234  & 0.0634  \\
    3275  & 14.9791  & 14.9937  & 0.0146  \\
    4433  & 14.6373  & 14.7432  & 0.1059  \\
    6149  & 14.4671  & 14.4476  & 0.0195  \\
    8030  & 14.2280  & 14.1836  & 0.0444  \\
    10121 & 13.8412  & 13.9349  & 0.0937  \\
    7892  & 14.0501  & 14.2014  & 0.1513  \\
    16469 & 13.0848  & 13.3354  & 0.2506  \\
    4604  & 14.9887  & 14.7104  & 0.2783  \\
    10175 & 13.7093  & 13.9289  & 0.2196  \\
    15853 & 13.2100  & 13.3868  & 0.1768  \\
    16511 & 13.0593  & 13.3320  & 0.2727  \\
    4900  & 14.8488  & 14.6557  & 0.1931  \\
    3108  & 14.8956  & 15.0350  & 0.1394  \\
    4235  & 14.9564  & 14.7824  & 0.1740  \\
    2812  & 15.1193  & 15.1125  & 0.0068  \\
    19929 & 12.8762  & 13.0648  & 0.1886  \\
    12419 & 13.6861  & 13.6969  & 0.0108  \\
    2781  & 15.2080  & 15.1209  & 0.0871  \\
    33790 & 12.0710  & 12.1618  & 0.0908  \\
    7255  & 14.3494  & 14.2867  & 0.0627  \\
    4759  & 14.6110  & 14.6815  & 0.0705  \\
    5817  & 14.5363  & 14.4998  & 0.0365  \\
    9010  & 14.1731  & 14.0623  & 0.1108  \\
    16075 & 13.3508  & 13.3682  & 0.0174  \\
    4209  & 15.0816  & 14.7876  & 0.2940  \\
    14374 & 13.4471  & 13.5150  & 0.0679  \\
    8362  & 14.1853  & 14.1415  & 0.0438  \\
    25959 & 12.4494  & 12.6456  & 0.1962  \\
    15994 & 13.5760  & 13.3749  & 0.2011  \\
    11748 & 13.7758  & 13.7634  & 0.0124  \\
    27398 & 12.6873  & 12.5526  & 0.1347  \\
    5653  & 14.3732  & 14.5263  & 0.1531  \\
    15143 & 13.4274  & 13.4474  & 0.0200  \\
    9316  & 13.8082  & 14.0263  & 0.2181  \\
    27634 & 12.7848  & 12.5375  & 0.2473  \\
    36697 & 12.4733  & 11.9934  & 0.4799  \\
    15403 & 13.5678  & 13.4250  & 0.1428  \\
    28387 & 12.6146  & 12.4899  & 0.1247  \\
    45666 & 11.4969  & 11.4969  & 0.0000  \\
    9441  & 14.0328  & 14.0117  & 0.0211  \\
    6378  & 14.1172  & 14.4127  & 0.2955  \\
    23743 & 13.0053  & 12.7935  & 0.2118  \\
    \hline
  \label{tab:addlabel}%
\end{longtable}%
\textbf{Note.} Column 1: sum of RGB $R_i+G_i+B_i$ from each star patch. Column 2: expected output of apparent magnitude. Column 3: measured magnitude. Column 4: absolute error of each measured magnitude.
\newline

\begin{longtable}{rrrrr}
 \caption{\small{Measurement of effective temperature with labelled data from the image centering HIP67301.}}
 \\
 \hline
 \multicolumn{1}{l}{$R_i$} & \multicolumn{1}{l}{$B_i$} & \multicolumn{1}{l}{${T_{eff}}_i$} & \multicolumn{1}{l}{$\hat{T}_{eff}$} & \multicolumn{1}{l}{$e_i$}\\
 \hline
 \endhead
    9125  & 10902 & 5798.75  & 5721.25  & 0.01  \\
    12822 & 15273 & 6342.75  & 5977.76  & 0.06  \\
    2499  & 2100  & 4364.00  & 4638.79  & 0.06  \\
    1950  & 1692  & 5740.67  & 4750.22  & 0.20  \\
    3992  & 4281  & 5790.50  & 5476.52  & 0.05  \\
    8918  & 9046  & 5544.00  & 5342.71  & 0.04  \\
    1750  & 2120  & 5367.00  & 5576.43  & 0.04  \\
    2852  & 3411  & 5248.10  & 5630.44  & 0.07  \\
    1758  & 2142  & 4956.66  & 5604.32  & 0.13  \\
    10261 & 12247 & 5875.00  & 5530.61  & 0.06  \\
    3060  & 3562  & 4897.33  & 5584.26  & 0.14  \\
    3599  & 4448  & 5770.25  & 5753.18  & 0.00  \\
    7988  & 9992  & 5865.33  & 5944.31  & 0.01  \\
    1698  & 2258  & 5298.00  & 5809.36  & 0.10  \\
    3376  & 4264  & 6502.25  & 5740.16  & 0.12  \\
    8804  & 10547 & 5867.00  & 5754.42  & 0.02  \\
    2386  & 2321  & 4844.41  & 5021.76  & 0.04  \\
    9544  & 11405 & 5853.50  & 5638.23  & 0.04  \\
    2308  & 2834  & 5403.25  & 5647.26  & 0.05  \\
    5749  & 7278  & 5779.33  & 5682.16  & 0.02  \\
    5886  & 6762  & 5029.00  & 5731.04  & 0.14  \\
    6216  & 6837  & 5766.00  & 5661.00  & 0.02  \\
    974   & 1198  & 4973.67  & 5520.02  & 0.11  \\
    1782  & 2414  & 6138.00  & 5829.75  & 0.05  \\
    4569  & 5201  & 5363.00  & 5693.06  & 0.06  \\
    8676  & 10890 & 6272.00  & 5774.97  & 0.08  \\
    2116  & 2595  & 4957.75  & 5644.18  & 0.14  \\
    904   & 1099  & 5449.83  & 5461.44  & 0.00  \\
    2061  & 2388  & 5322.00  & 5479.82  & 0.03  \\
    1393  & 1362  & 5454.44  & 5134.83  & 0.06  \\
    6760  & 7305  & 5084.58  & 5704.90  & 0.12  \\
    6173  & 7793  & 5556.00  & 5760.21  & 0.04  \\
    3277  & 3619  & 5830.25  & 5437.34  & 0.07  \\
    6155  & 7662  & 5286.90  & 5794.34  & 0.10  \\
    4285  & 4349  & 5515.94  & 5419.68  & 0.02  \\
    4275  & 4392  & 5217.08  & 5475.33  & 0.05  \\
    2201  & 1502  & 4115.33  & 4248.81  & 0.03  \\
    8377  & 9008  & 5025.33  & 5637.60  & 0.12  \\
    898   & 1234  & 5332.00  & 5651.24  & 0.06  \\
    1680  & 2057  & 5710.75  & 5588.75  & 0.02  \\
    12620 & 15025 & 4986.31  & 6446.55  & 0.29  \\
    1328  & 1535  & 6171.61  & 5423.83  & 0.12  \\
    3566  & 4262  & 5811.75  & 5674.30  & 0.02  \\
    3968  & 4762  & 5112.29  & 5750.94  & 0.12  \\
    1574  & 1953  & 5880.67  & 5601.72  & 0.05  \\
    15600 & 17538 & 5037.70  & 5738.42  & 0.14  \\
    2590  & 3013  & 5663.00  & 5513.45  & 0.03  \\
    3405  & 3700  & 5245.11  & 5429.55  & 0.04  \\
    640   & 625   & 5037.70  & 5238.14  & 0.04  \\
    5221  & 6161  & 5810.25  & 5687.12  & 0.02  \\
    3638  & 4553  & 5836.00  & 5774.12  & 0.01  \\
    1805  & 2266  & 5335.11  & 5672.28  & 0.06  \\
    965   & 1144  & 5045.10  & 5459.85  & 0.08  \\
    3674  & 4431  & 5297.00  & 5731.00  & 0.08  \\
    794   & 1031  & 5601.65  & 5510.02  & 0.02  \\
    1247  & 1315  & 4912.80  & 5291.36  & 0.08  \\
    5401  & 6708  & 6297.00  & 5617.60  & 0.11  \\
    731   & 928   & 5971.15  & 5401.62  & 0.10  \\
    1283  & 1429  & 5635.25  & 5366.69  & 0.05  \\
    4171  & 5166  & 5786.00  & 5756.89  & 0.01  \\
    2358  & 2783  & 5805.25  & 5531.08  & 0.05  \\
    999   & 1452  & 5564.05  & 5785.92  & 0.04  \\
    10623 & 12219 & 5005.00  & 5624.99  & 0.12  \\
    14029 & 17051 & 6806.22  & 5804.88  & 0.15  \\
    1824  & 2013  & 4910.83  & 5369.84  & 0.09  \\
    4486  & 4620  & 5813.67  & 5477.84  & 0.06  \\
    2604  & 2963  & 5745.00  & 5446.51  & 0.05  \\
    2405  & 2333  & 5045.92  & 5007.53  & 0.01  \\
    15443 & 18055 & 5933.33  & 5394.39  & 0.09  \\
    2668  & 3236  & 5355.75  & 5648.02  & 0.05  \\
    1981  & 1481  & 4227.50  & 4513.31  & 0.07  \\
    1453  & 995   & 4190.64  & 4653.30  & 0.11  \\
    1569  & 1700  & 5502.91  & 5317.63  & 0.03  \\
    6856  & 7486  & 6138.00  & 5609.70  & 0.09  \\
    1444  & 1295  & 5312.00  & 4968.40  & 0.06  \\
    11578 & 12896 & 5783.25  & 5511.49  & 0.05  \\
    3766  & 3654  & 5811.25  & 5128.70  & 0.12  \\
    3881  & 3194  & 4947.05  & 4637.99  & 0.06  \\
    4150  & 4975  & 5831.33  & 5720.07  & 0.02  \\
    12323 & 12643 & 6362.00  & 5035.64  & 0.21  \\
    2487  & 3472  & 5925.30  & 5995.68  & 0.01  \\
    6809  & 6597  & 5803.00  & 5524.76  & 0.05  \\
    2438  & 2447  & 5347.75  & 5096.40  & 0.05  \\
    1669  & 1835  & 5112.47  & 5355.87  & 0.05  \\
    14903 & 14790 & 4999.64  & 5066.22  & 0.01  \\
    13002 & 14096 & 5858.00  & 5677.54  & 0.03  \\
    2843  & 3640  & 5782.75  & 5795.63  & 0.00  \\
    1476  & 1697  & 5515.04  & 5439.51  & 0.01  \\
    4975  & 5749  & 5004.00  & 5749.07  & 0.15  \\
    1811  & 1785  & 4911.89  & 5108.09  & 0.04  \\
    2674  & 2845  & 5136.00  & 5275.72  & 0.03  \\
    2673  & 2559  & 4344.79  & 4992.13  & 0.15  \\
    1496  & 1764  & 6451.00  & 5468.42  & 0.15  \\
    2372  & 2961  & 5479.50  & 5699.96  & 0.04  \\
    1516  & 1780  & 5087.00  & 5496.81  & 0.08  \\
    3203  & 3323  & 5676.00  & 5241.62  & 0.08  \\
    9957  & 10325 & 4898.66  & 5405.12  & 0.10  \\
    3693  & 4033  & 6256.00  & 5457.14  & 0.13  \\
    8665  & 8181  & 5626.00  & 5163.13  & 0.08  \\
    9652  & 10556 & 5777.25  & 5392.22  & 0.07  \\
    2454  & 2628  & 5440.24  & 5274.95  & 0.03  \\
    3189  & 3038  & 4665.58  & 5011.02  & 0.07  \\
    8502  & 8315  & 5408.33  & 5364.43  & 0.01  \\
    2092  & 2116  & 5144.33  & 5132.20  & 0.00  \\
    3146  & 3525  & 5568.28  & 5470.46  & 0.02  \\
    6303  & 6298  & 5131.81  & 5723.49  & 0.12  \\
    4688  & 4208  & 5409.00  & 5186.05  & 0.04  \\
    7723  & 8678  & 5120.75  & 5760.50  & 0.12  \\
    1233  & 1709  & 6395.75  & 5747.58  & 0.10  \\
    2241  & 1980  & 4458.00  & 4799.22  & 0.08  \\
    6431  & 6691  & 5755.00  & 5623.39  & 0.02  \\
    3718  & 4674  & 5480.29  & 5812.10  & 0.06  \\
    3710  & 4175  & 5094.67  & 5582.66  & 0.10  \\
    1632  & 2232  & 6752.50  & 5791.99  & 0.14  \\
    2353  & 2703  & 5421.33  & 5467.72  & 0.01  \\
    1220  & 1485  & 5012.04  & 5540.56  & 0.11  \\
    1465  & 1678  & 5992.11  & 5420.02  & 0.10  \\
    2251  & 2448  & 5294.05  & 5318.13  & 0.00  \\
    12130 & 12970 & 5575.18  & 5518.42  & 0.01  \\
    7110  & 8816  & 6322.67  & 5705.58  & 0.10  \\
    12670 & 13085 & 4945.94  & 5727.75  & 0.16  \\
    3030  & 3177  & 4938.30  & 5278.67  & 0.07  \\
    2923  & 2920  & 5148.85  & 5105.36  & 0.01  \\
    2442  & 3102  & 6197.00  & 5723.94  & 0.08  \\
    11442 & 13902 & 6321.00  & 5388.74  & 0.15  \\
    5530  & 6753  & 6342.67  & 5617.21  & 0.11  \\
    5100  & 5683  & 6214.00  & 5626.46  & 0.09  \\
    3213  & 3014  & 5752.00  & 4914.45  & 0.15  \\
    9727  & 11072 & 4783.59  & 5646.44  & 0.18  \\
    10807 & 12169 & 5015.69  & 5563.43  & 0.11  \\
    2502  & 2205  & 3914.55  & 4791.16  & 0.22  \\
    4007  & 4569  & 5115.33  & 5649.70  & 0.10  \\
    2557  & 3213  & 5939.75  & 5720.62  & 0.04  \\
    823   & 1134  & 5342.33  & 5626.37  & 0.05  \\
    1244  & 1295  & 6808.00  & 5203.39  & 0.24  \\
    2127  & 2051  & 5485.00  & 4999.34  & 0.09  \\
    2794  & 3663  & 5934.50  & 5848.22  & 0.01  \\
    7350  & 8097  & 5783.50  & 5644.71  & 0.02  \\
    1480  & 1986  & 5760.88  & 5766.89  & 0.00  \\
    1853  & 2556  & 6824.50  & 5843.81  & 0.14  \\
    1126  & 1387  & 5344.50  & 5533.68  & 0.04  \\
    1116  & 1510  & 4633.64  & 5732.42  & 0.24  \\
    9054  & 9625  & 5889.00  & 5389.13  & 0.08  \\
    3438  & 4073  & 5811.75  & 5645.95  & 0.03  \\
    \hline
  \label{tab:addlabel}%
\end{longtable}%
\textbf{Note.} Column 1: the R channel reading from each star patch. Column 2: the B channel reading from each star patch. Column 3: expected output of effective temperature. Column 4: measured effective temperature. Column 5: generalization error tested on each labelled data.
\newline

\begin{longtable}{rrrrr}
  \caption{\small{Measurement of effective temperature using unknown data in HIP65474 with the $\hat{T}_{eff}$ function obtained from data in HIP67301.}}
  \\
  \hline
  \multicolumn{1}{l}{$R$} & \multicolumn{1}{l}{$B$} & \multicolumn{1}{l}{${T_eff}_i$} & \multicolumn{1}{l}{$\hat{T}_{eff}$} & \multicolumn{1}{l}{$e_i$} \\
  \hline
  \endhead
    8403  & 8730  & 6027.00  & 5490.99  & 0.09  \\
    5282  & 5277  & 6732.10  & 5581.39  & 0.17  \\
    2100  & 1514  & 5104.50  & 4376.25  & 0.14  \\
    1092  & 1127  & 5437.67  & 5251.24  & 0.03  \\
    3521  & 3018  & 4915.20  & 4733.16  & 0.04  \\
    1694  & 1760  & 5982.50  & 5226.25  & 0.13  \\
    5256  & 3003  & 4848.37  & 4140.71  & 0.15  \\
    2718  & 1965  & 3993.00  & 4204.19  & 0.05  \\
    1407  & 1439  & 4945.00  & 5219.45  & 0.06  \\
    6247  & 6148  & 5946.50  & 5615.23  & 0.06  \\
    4991  & 3961  & 4590.51  & 5016.51  & 0.09  \\
    1390  & 1325  & 5055.77  & 5100.24  & 0.01  \\
    8263  & 7216  & 5082.69  & 5284.54  & 0.04  \\
    11602 & 8565  & 4319.95  & 4286.58  & 0.01  \\
    2059  & 2079  & 4973.00  & 5131.51  & 0.03  \\
    3809  & 3669  & 4947.87  & 5166.48  & 0.04  \\
    5964  & 5211  & 4882.15  & 5489.41  & 0.12  \\
    5013  & 5712  & 6506.09  & 5691.02  & 0.13  \\
    1088  & 856   & 4937.00  & 4909.23  & 0.01  \\
    1890  & 1886  & 6044.50  & 5119.51  & 0.15  \\
    2286  & 2577  & 5593.85  & 5411.00  & 0.03  \\
    5413  & 5459  & 5483.25  & 5605.81  & 0.02  \\
    8597  & 8409  & 5761.00  & 5355.93  & 0.07  \\
    3782  & 4169  & 6310.00  & 5523.87  & 0.12  \\
    2330  & 1830  & 4857.14  & 4488.02  & 0.08  \\
    8517  & 6942  & 4971.09  & 5161.12  & 0.04  \\
    927   & 878   & 4947.62  & 5150.13  & 0.04  \\
    676   & 716   & 6449.00  & 5256.62  & 0.18  \\
    2907  & 2972  & 5879.00  & 5173.63  & 0.12  \\
    1603  & 1910  & 5382.75  & 5521.87  & 0.03  \\
    2165  & 2450  & 6603.50  & 5417.50  & 0.18  \\
    1360  & 1225  & 4943.00  & 5014.62  & 0.01  \\
    2685  & 2602  & 5467.02  & 5002.81  & 0.08  \\
    989   & 989   & 6541.00  & 5211.55  & 0.20  \\
    1764  & 1645  & 4969.00  & 4992.54  & 0.00  \\
    7328  & 8200  & 6554.00  & 5677.74  & 0.13  \\
    9589  & 7740  & 4452.64  & 4841.75  & 0.09  \\
    5318  & 4707  & 4830.40  & 5389.68  & 0.12  \\
    2987  & 3426  & 6744.00  & 5513.63  & 0.18  \\
    1499  & 1829  & 5427.00  & 5564.18  & 0.03  \\
    1683  & 1581  & 5004.00  & 5022.22  & 0.00  \\
    1165  & 1272  & 4980.33  & 5336.22  & 0.07  \\
    4384  & 4724  & 6413.00  & 5572.65  & 0.13  \\
    1820  & 1692  & 5759.66  & 4975.75  & 0.14  \\
    659   & 787   & 6020.40  & 5367.92  & 0.11  \\
    14611 & 14944 & 5718.60  & 5295.83  & 0.07  \\
    3661  & 3652  & 5854.00  & 5231.87  & 0.11  \\
    5323  & 5112  & 4949.00  & 5532.60  & 0.12  \\
    9937  & 10569 & 6318.50  & 5344.35  & 0.15  \\
    15525 & 17697 & 6302.00  & 5497.72  & 0.13  \\
    1940  & 1648  & 4428.74  & 4760.89  & 0.07  \\
    7049  & 6429  & 5085.45  & 5523.77  & 0.09  \\
    1497  & 1748  & 5796.00  & 5472.52  & 0.06  \\
    626   & 767   & 6045.33  & 5381.67  & 0.11  \\
    6514  & 6927  & 5965.67  & 5645.91  & 0.05  \\
    861   & 790   & 5327.00  & 5121.71  & 0.04  \\
    1464  & 1151  & 5428.33  & 4776.38  & 0.12  \\
    2938  & 2334  & 5627.97  & 4436.84  & 0.21  \\
    4801  & 4445  & 6043.75  & 5347.63  & 0.12  \\
    3370  & 2904  & 5381.95  & 4719.28  & 0.12  \\
    3024  & 2743  & 5949.50  & 4829.13  & 0.19  \\
    2775  & 2318  & 5261.00  & 4582.53  & 0.13  \\
    2689  & 2355  & 5773.67  & 4717.36  & 0.18  \\
    1092  & 1157  & 5770.00  & 5288.99  & 0.08  \\
    3841  & 3945  & 5954.00  & 5355.28  & 0.10  \\
    675   & 766   & 4429.50  & 5321.92  & 0.20  \\
    3092  & 2758  & 5044.00  & 4785.38  & 0.05  \\
    2214  & 2290  & 5939.61  & 5183.56  & 0.13  \\
    1401  & 1236  & 6302.06  & 4972.50  & 0.21  \\
    1396  & 1340  & 5869.00  & 5110.94  & 0.13  \\
    1374  & 1457  & 5744.00  & 5286.02  & 0.08  \\
    1058  & 1420  & 8427.67  & 5655.79  & 0.33  \\
    800   & 1003  & 6009.67  & 5469.40  & 0.09  \\
    5903  & 6213  & 5478.33  & 5654.75  & 0.03  \\
    1267  & 1478  & 8657.00  & 5454.03  & 0.37  \\
    1940  & 2186  & 5836.00  & 5403.12  & 0.07  \\
    675   & 875   & 5572.01  & 5460.40  & 0.02  \\
    289   & 352   & 4909.00  & 5222.62  & 0.06  \\
    2631  & 2617  & 5836.00  & 5075.40  & 0.13  \\
    654   & 684   & 5740.00  & 5241.97  & 0.09  \\
    2561  & 2093  & 5802.00  & 4542.30  & 0.22  \\
    1938  & 1947  & 5359.33  & 5130.04  & 0.04  \\
    625   & 877   & 5824.00  & 5522.10  & 0.05  \\
    449   & 351   & 5409.00  & 5048.24  & 0.07  \\
    1128  & 1390  & 5767.50  & 5529.55  & 0.04  \\
    1609  & 1250  & 4917.31  & 4703.27  & 0.04  \\
    4510  & 4150  & 5354.00  & 5245.60  & 0.02  \\
    3515  & 3164  & 4847.78  & 4885.49  & 0.01  \\
    1788  & 1886  & 5766.00  & 5252.61  & 0.09  \\
    2775  & 2493  & 5192.20  & 4788.02  & 0.08  \\
    499   & 679   & 5124.93  & 5416.22  & 0.06  \\
    1232  & 1291  & 5650.65  & 5270.78  & 0.07  \\
    675   & 821   & 5919.33  & 5392.02  & 0.09  \\
    3311  & 3006  & 5109.00  & 4870.78  & 0.05  \\
    1967  & 2151  & 5205.50  & 5329.93  & 0.02  \\
    1352  & 1411  & 5607.74  & 5258.86  & 0.06  \\
    1737  & 1558  & 4922.00  & 4920.50  & 0.00  \\
    10495 & 10100 & 4954.00  & 5036.17  & 0.02  \\
    4261  & 3909  & 5367.00  & 5158.28  & 0.04  \\
    1019  & 947   & 4989.25  & 5118.57  & 0.03  \\
    3749  & 3818  & 5794.25  & 5311.11  & 0.08  \\
    4329  & 3901  & 5765.67  & 5129.66  & 0.11  \\
    6486  & 7031  & 6401.25  & 5656.07  & 0.12  \\
    2999  & 2770  & 6014.67  & 4881.00  & 0.19  \\
    2282  & 1974  & 5761.33  & 4727.50  & 0.18  \\
    2230  & 1796  & 4948.32  & 4570.32  & 0.08  \\
    1762  & 1787  & 6383.00  & 5168.67  & 0.19  \\
    1859  & 2304  & 5436.50  & 5632.46  & 0.04  \\
    1842  & 1500  & 4601.09  & 4704.73  & 0.02  \\
    1171  & 1258  & 5776.49  & 5310.87  & 0.08  \\
    3123  & 2392  & 5552.00  & 4333.54  & 0.22  \\
    986   & 1070  & 4946.25  & 5318.05  & 0.08  \\
    1982  & 1926  & 5781.00  & 5048.24  & 0.13  \\
    1120  & 1461  & 5373.00  & 5625.33  & 0.05  \\
    4590  & 3745  & 4994.33  & 4934.67  & 0.01  \\
    9023  & 9762  & 6833.00  & 5499.02  & 0.20  \\
    1631  & 1564  & 4908.33  & 5071.39  & 0.03  \\
    2122  & 1934  & 5331.25  & 4878.80  & 0.08  \\
    1336  & 1086  & 4270.00  & 4869.30  & 0.14  \\
    4554  & 3755  & 5034.67  & 4951.93  & 0.02  \\
    4165  & 4589  & 5794.00  & 5582.74  & 0.04  \\
    812   & 753   & 5131.38  & 5136.18  & 0.00  \\
    9825  & 9688  & 5975.50  & 5163.08  & 0.14  \\
    1344  & 1187  & 4866.18  & 4988.06  & 0.03  \\
    703   & 722   & 5083.52  & 5231.54  & 0.03  \\
    4014  & 4083  & 5686.80  & 5369.89  & 0.06  \\
    1765  & 1584  & 6250.00  & 4915.23  & 0.21  \\
    905   & 669   & 5122.04  & 4907.79  & 0.04  \\
    12765 & 10015 & 5265.84  & 4526.44  & 0.14  \\
    3131  & 3052  & 5750.00  & 5059.17  & 0.12  \\
    1770  & 1605  & 5362.00  & 4934.74  & 0.08  \\
    3630  & 3240  & 5687.00  & 4886.75  & 0.14  \\
    11642 & 12449 & 6181.00  & 5447.49  & 0.12  \\
    1305  & 1338  & 5744.00  & 5231.47  & 0.09  \\
    11897 & 11055 & 5084.20  & 5021.08  & 0.01  \\
    11885 & 10959 & 4895.75  & 4992.18  & 0.02  \\
    12011 & 9249  & 4086.18  & 4400.59  & 0.08  \\
    1814  & 2107  & 5379.50  & 5473.68  & 0.02  \\
    1477  & 1478  & 6130.00  & 5173.19  & 0.16  \\
    945   & 1223  & 5782.00  & 5560.75  & 0.04  \\
    1492  & 1462  & 5343.55  & 5133.04  & 0.04  \\
    2054  & 2111  & 5431.00  & 5175.03  & 0.05  \\
    2614  & 2683  & 5829.00  & 5163.78  & 0.11  \\
    3328  & 3400  & 5808.50  & 5232.51  & 0.10  \\
    2866  & 2323  & 4557.84  & 4495.14  & 0.01  \\
    5333  & 5726  & 5880.50  & 5657.67  & 0.04  \\
    1350  & 1742  & 5740.90  & 5657.71  & 0.01  \\
    3333  & 3326  & 5289.67  & 5163.54  & 0.02  \\
    5353  & 5407  & 6423.67  & 5602.08  & 0.13  \\
    5690  & 5335  & 4926.57  & 5549.30  & 0.13  \\
    1773  & 1637  & 5353.53  & 4970.54  & 0.07  \\
    967   & 1108  & 5684.75  & 5390.26  & 0.05  \\
    1556  & 1435  & 5855.09  & 5012.28  & 0.14  \\
    876   & 984   & 5766.00  & 5350.44  & 0.07  \\
    6119  & 7066  & 5827.75  & 5685.39  & 0.02  \\
    4024  & 4271  & 5849.00  & 5477.49  & 0.06  \\
    827   & 975   & 5415.00  & 5400.69  & 0.00  \\
    11017 & 11404 & 5520.07  & 5261.39  & 0.05  \\
    2152  & 2584  & 6373.00  & 5572.08  & 0.13  \\
    1633  & 1598  & 5313.00  & 5110.69  & 0.04  \\
    1769  & 2105  & 5821.50  & 5528.65  & 0.05  \\
    2796  & 3157  & 6905.00  & 5449.30  & 0.21  \\
    5230  & 5455  & 5112.00  & 5626.01  & 0.10  \\
    1619  & 1399  & 4109.80  & 4880.62  & 0.19  \\
    4664  & 4901  & 5407.00  & 5570.81  & 0.03  \\
    2813  & 2912  & 6188.33  & 5201.19  & 0.16  \\
    7998  & 9200  & 6269.00  & 5726.77  & 0.09  \\
    6141  & 4927  & 5432.00  & 5385.93  & 0.01  \\
    4270  & 3440  & 5840.00  & 4769.01  & 0.18  \\
    9898  & 8660  & 6168.67  & 4892.23  & 0.21  \\
    1842  & 1713  & 5383.33  & 4972.37  & 0.08  \\
    5936  & 4451  & 5566.00  & 5211.04  & 0.06  \\
    3715  & 2311  & 4314.12  & 3777.41  & 0.12  \\
    9294  & 8868  & 6391.67  & 5179.89  & 0.19  \\
    12820 & 11803 & 5893.67  & 5132.07  & 0.13  \\
    5744  & 4853  & 5482.19  & 5400.27  & 0.01  \\
    12388 & 7417  & 4677.98  & 3676.24  & 0.21  \\
    14316 & 16330 & 6365.71  & 5972.44  & 0.06  \\
    2967  & 3430  & 4982.69  & 5533.17  & 0.11  \\
    2230  & 1990  & 5372.33  & 4811.44  & 0.10  \\
    8311  & 7808  & 5520.07  & 5350.83  & 0.03  \\

    \hline
  \label{tab:addlabel}%
\end{longtable}%
\textbf{Note.} Column 1: R channel reading from each star patch in the image centering HIP65474. Column 2: B channel reading. Column 3: expected output of effective temperature. Column 4: measured effective temperature with $\hat{T}_{eff}$ function from labelled data in HIp67301. Column 5: generalization error of $\hat{T}_{eff}$. 
\newline 

\normalsize
\textbf{Reference}
\scriptsize
\newline

\setlength{\parindent}{0em}

Brown, A.G.A., et al., 2018, A\&A, 616, A1
\newline
Carroll, W.B.; Ostlie, D.A., 2014, "An Introduction to Modern Astrophysics, Second Edition" (2014, Pearson)
\newline
Craven, P.; G. Wahba, 1979, NM, 31, 377
\newline
Johnson, H.L.; Morgan, W.W., 1953, ApJ, 117, 313
\newline
Karttunen, H., et al., 2017, "Fundamental Astronomy, Sixth Edition" (2017, Springer)
\newline
Phillips, K.J.H., 1995, "Guide to the sun", Cambridge University Press
\newline
Poggio, T.; F. Girosi,1990, Sci, 247, 978
\newline
Tikhonov, A.N., 1963, SM, 4, 1035

\end{document}